\RequirePackage{lineno}

\documentclass[aps,prd,twocolumn,showpacs,superscriptaddress,groupedaddress,nofootinbib]{revtex4} 

\usepackage{graphicx}
\usepackage{amssymb}
\usepackage{amsmath}
\usepackage{verbatim}
\usepackage{subfigure}
\usepackage{enumerate}
\usepackage{mathtools}
\usepackage{multirow}
\usepackage{alphalph}

\newcommand{\bl}{$\text{MVA}_{bl}$}

\newcommand{\syst}{\textit{System8}}
\newcommand{\msv}{M$_{\text{SV}}$}

\newcommand{\dzero}{D0}
\newcommand{\Dzero}{D0}

\newcommand{\pt}{$p_{T}$~}
\newcommand{\pte}{$p_{T}$}

\usepackage{color}

\begin{document}
\hspace{5.2in} \mbox{FERMILAB-PUB-13-579-E}
\title{Improved \textit{b} quark jet identification at the \Dzero\ experiment}

\affiliation{LAFEX, Centro Brasileiro de Pesquisas F\'{i}sicas, Rio de Janeiro, Brazil}
\affiliation{Universidade do Estado do Rio de Janeiro, Rio de Janeiro, Brazil}
\affiliation{Universidade Federal do ABC, Santo Andr\'e, Brazil}
\affiliation{University of Science and Technology of China, Hefei, People's Republic of China}
\affiliation{Universidad de los Andes, Bogot\'a, Colombia}
\affiliation{Charles University, Faculty of Mathematics and Physics, Center for Particle Physics, Prague, Czech Republic}
\affiliation{Czech Technical University in Prague, Prague, Czech Republic}
\affiliation{Institute of Physics, Academy of Sciences of the Czech Republic, Prague, Czech Republic}
\affiliation{Universidad San Francisco de Quito, Quito, Ecuador}
\affiliation{LPC, Universit\'e Blaise Pascal, CNRS/IN2P3, Clermont, France}
\affiliation{LPSC, Universit\'e Joseph Fourier Grenoble 1, CNRS/IN2P3, Institut National Polytechnique de Grenoble, Grenoble, France}
\affiliation{CPPM, Aix-Marseille Universit\'e, CNRS/IN2P3, Marseille, France}
\affiliation{LAL, Universit\'e Paris-Sud, CNRS/IN2P3, Orsay, France}
\affiliation{LPNHE, Universit\'es Paris VI and VII, CNRS/IN2P3, Paris, France}
\affiliation{CEA, Irfu, SPP, Saclay, France}
\affiliation{IPHC, Universit\'e de Strasbourg, CNRS/IN2P3, Strasbourg, France}
\affiliation{IPNL, Universit\'e Lyon 1, CNRS/IN2P3, Villeurbanne, France and Universit\'e de Lyon, Lyon, France}
\affiliation{III. Physikalisches Institut A, RWTH Aachen University, Aachen, Germany}
\affiliation{Physikalisches Institut, Universit\"at Freiburg, Freiburg, Germany}
\affiliation{II. Physikalisches Institut, Georg-August-Universit\"at G\"ottingen, G\"ottingen, Germany}
\affiliation{Institut f\"ur Physik, Universit\"at Mainz, Mainz, Germany}
\affiliation{Ludwig-Maximilians-Universit\"at M\"unchen, M\"unchen, Germany}
\affiliation{Panjab University, Chandigarh, India}
\affiliation{Delhi University, Delhi, India}
\affiliation{Tata Institute of Fundamental Research, Mumbai, India}
\affiliation{University College Dublin, Dublin, Ireland}
\affiliation{Korea Detector Laboratory, Korea University, Seoul, Korea}
\affiliation{CINVESTAV, Mexico City, Mexico}
\affiliation{Nikhef, Science Park, Amsterdam, the Netherlands}
\affiliation{Radboud University Nijmegen, Nijmegen, the Netherlands}
\affiliation{Joint Institute for Nuclear Research, Dubna, Russia}
\affiliation{Institute for Theoretical and Experimental Physics, Moscow, Russia}
\affiliation{Moscow State University, Moscow, Russia}
\affiliation{Institute for High Energy Physics, Protvino, Russia}
\affiliation{Petersburg Nuclear Physics Institute, St. Petersburg, Russia}
\affiliation{Instituci\'{o} Catalana de Recerca i Estudis Avan\c{c}ats (ICREA) and Institut de F\'{i}sica d'Altes Energies (IFAE), Barcelona, Spain}
\affiliation{Uppsala University, Uppsala, Sweden}
\affiliation{Taras Shevchenko National University of Kyiv, Kiev, Ukraine}
\affiliation{Lancaster University, Lancaster LA1 4YB, United Kingdom}
\affiliation{Imperial College London, London SW7 2AZ, United Kingdom}
\affiliation{The University of Manchester, Manchester M13 9PL, United Kingdom}
\affiliation{University of Arizona, Tucson, Arizona 85721, USA}
\affiliation{University of California Riverside, Riverside, California 92521, USA}
\affiliation{Florida State University, Tallahassee, Florida 32306, USA}
\affiliation{Fermi National Accelerator Laboratory, Batavia, Illinois 60510, USA}
\affiliation{University of Illinois at Chicago, Chicago, Illinois 60607, USA}
\affiliation{Northern Illinois University, DeKalb, Illinois 60115, USA}
\affiliation{Northwestern University, Evanston, Illinois 60208, USA}
\affiliation{Indiana University, Bloomington, Indiana 47405, USA}
\affiliation{Purdue University Calumet, Hammond, Indiana 46323, USA}
\affiliation{University of Notre Dame, Notre Dame, Indiana 46556, USA}
\affiliation{Iowa State University, Ames, Iowa 50011, USA}
\affiliation{University of Kansas, Lawrence, Kansas 66045, USA}
\affiliation{Louisiana Tech University, Ruston, Louisiana 71272, USA}
\affiliation{Northeastern University, Boston, Massachusetts 02115, USA}
\affiliation{University of Michigan, Ann Arbor, Michigan 48109, USA}
\affiliation{Michigan State University, East Lansing, Michigan 48824, USA}
\affiliation{University of Mississippi, University, Mississippi 38677, USA}
\affiliation{University of Nebraska, Lincoln, Nebraska 68588, USA}
\affiliation{Rutgers University, Piscataway, New Jersey 08855, USA}
\affiliation{Princeton University, Princeton, New Jersey 08544, USA}
\affiliation{State University of New York, Buffalo, New York 14260, USA}
\affiliation{University of Rochester, Rochester, New York 14627, USA}
\affiliation{State University of New York, Stony Brook, New York 11794, USA}
\affiliation{Brookhaven National Laboratory, Upton, New York 11973, USA}
\affiliation{Langston University, Langston, Oklahoma 73050, USA}
\affiliation{University of Oklahoma, Norman, Oklahoma 73019, USA}
\affiliation{Oklahoma State University, Stillwater, Oklahoma 74078, USA}
\affiliation{Brown University, Providence, Rhode Island 02912, USA}
\affiliation{University of Texas, Arlington, Texas 76019, USA}
\affiliation{Southern Methodist University, Dallas, Texas 75275, USA}
\affiliation{Rice University, Houston, Texas 77005, USA}
\affiliation{University of Virginia, Charlottesville, Virginia 22904, USA}
\affiliation{University of Washington, Seattle, Washington 98195, USA}
\author{V.M.~Abazov} \affiliation{Joint Institute for Nuclear Research, Dubna, Russia}
\author{B.~Abbott} \affiliation{University of Oklahoma, Norman, Oklahoma 73019, USA}
\author{B.S.~Acharya} \affiliation{Tata Institute of Fundamental Research, Mumbai, India}
\author{M.~Adams} \affiliation{University of Illinois at Chicago, Chicago, Illinois 60607, USA}
\author{T.~Adams} \affiliation{Florida State University, Tallahassee, Florida 32306, USA}
\author{J.P.~Agnew} \affiliation{The University of Manchester, Manchester M13 9PL, United Kingdom}
\author{G.D.~Alexeev} \affiliation{Joint Institute for Nuclear Research, Dubna, Russia}
\author{G.~Alkhazov} \affiliation{Petersburg Nuclear Physics Institute, St. Petersburg, Russia}
\author{A.~Alton$^{a}$} \affiliation{University of Michigan, Ann Arbor, Michigan 48109, USA}
\author{A.~Askew} \affiliation{Florida State University, Tallahassee, Florida 32306, USA}
% ...
%
%
\collaboration{The D0 Collaboration\footnote{with visitors from
$^{a}$Augustana College, Sioux Falls, SD, USA,
$^{b}$The University of Liverpool, Liverpool, UK,
$^{c}$DESY, Hamburg, Germany,
$^{d}$Universidad Michoacana de San Nicolas de Hidalgo, Morelia, Mexico
$^{e}$SLAC, Menlo Park, CA, USA,
$^{f}$University College London, London, UK,
$^{g}$Centro de Investigacion en Computacion - IPN, Mexico City, Mexico,
$^{h}$Universidade Estadual Paulista, S\~ao Paulo, Brazil,
$^{i}$Karlsruher Institut f\"ur Technologie (KIT) - Steinbuch Centre for Computing (SCC),
D-76128 Karlsrue, Germany,
$^{j}$Office of Science, U.S. Department of Energy, Washington, D.C. 20585, USA,
$^{k}$American Association for the Advancement of Science, Washington, D.C. 20005, USA
and
$^{l}$Kiev Institute for Nuclear Research, Kiev, Ukraine
}} \noaffiliation
\vskip 0.25cm

\date{December 29, 2013}

 \begin{abstract}

The ability to identify jets which originated from $b$ quarks
is an important tool of the physics program of the D0 experiment 
at the Fermilab Tevatron $p\bar{p}$~collider. 
This article describes a new algorithm designed
to select jets originating from $b$ quarks while suppressing the
contamination caused by jets from other quark flavors and gluons. 
Additionally, a new technique, the SystemN method, for determining the misidentification
rate directly from data is presented. 

  \end{abstract}
\maketitle

\section{Introduction}
\label{sec:intro}

The identification of heavy flavor jets, in particular those originating from $b$ or $c$ quarks, is an important technique in 
particle physics and crucial for studies of top quark, the Higgs boson, and other rare processes~\cite{bid_nim, cdfhobit, cmsbtag}. 
The $b$ quark is significantly more massive, $m_{b}\approx5$~GeV, than the other quarks with 
the exception of the top quark. This, along with the long lifetimes of $b$ hadrons, is used 
to create algorithms for identifying jets which originate from $b$ quarks, 
called $b$ jets. 
These algorithms are of primary importance for many measurements and searches performed using 
the full \Dzero~Run II dataset, recorded from April 2002 until September 2011, with an integrated luminosity of 10 fb$^{-1}$.
This paper describes improvements in the D0 $b$ jet identification algorithm beyond those 
presented in Ref.\thinspace\cite{bid_nim} and a data-driven method for determining the misidentification
rates of the algorithms, that utilizes a new template-fitting method to extract the sample composition 
directly from the data. 

\section{The upgraded \Dzero\ detector}
\label{sec:d0upgrade}

The D0 detector is a general purpose hadron collider detector composed of a
tracking system, liquid-argon sampling calorimeter, and muon system~\cite{run2det}.
The central tracking system consists of a 
silicon microstrip tracker (SMT)~\cite{Ahmed:2010fx} and a central fiber tracker (CFT), 
both located within a 1.9~T superconducting solenoidal magnet, with
designs optimized for tracking and vertexing at pseudorapidities\footnote{D0 uses a right-handed coordinate system with the origin at the nominal collision point 
in the center of the detector. The direction of the proton 
beam is the $+z$ axis, and the $+y$ axis points vertically upwards. The 
polar angle, $\theta$, is defined such that $\theta=0$ is in the $+z$ direction. 
Pseudorapidity is defined as $\eta = - \ln(\tan \frac{\theta}{2})$. 
The azimuthal angle $\varphi$ is defined relative to the $x$ axis in the 
plane transverse to the proton beam direction.
The momentum of all particles is measured transverse to the beam direction, \pte.}
 $|\eta|<3$ and $|\eta|<2.5$, respectively. 
The tracking system enables an accurate measurement of a track's impact parameter (IP), 
i.e. the distance of closest approach of a track to the $p\bar{p}$ interaction vertex.

The calorimetry comprises a liquid-argon and uranium calorimeter, with a central section (CC) 
covering pseudorapidities $|\eta|\lesssim1.1$ and two forward sections (EC) extending the coverage
to $|\eta|\approx4.2$~\cite{calopaper}.
The muon system, covering $|\eta|<2$, consists of three layers of tracking detectors and scintillation trigger 
counters. One layer is located in front of 1.8~T magnetized iron toroids, and two are positioned after the toroids.
The luminosity is measured using plastic scintillator arrays located in front of the EC cryostats~\cite{lumi}. 
      
\section{Data and simulated samples}

The Run II data sample is broken into four subsamples
based on different beam and detector conditions. All figures
and numbers presented within this article will, for conciseness, be from the largest of the four periods,
corresponding to the final 4.4 fb$^{-1}$ of integrated luminosity recorded by the D0 detector. 
The data are selected by triggering on events containing at least two 
jets.

To simulate these events we use the {\sc pythia}~\cite{pythia} Monte Carlo (MC) event 
generator to create a large sample of multijet events.
These events contain jets originating from all types of partons. 
The fragmentation and decay of particles containing $b$ or $c$ quarks 
is modeled with {\sc evtgen}~\cite{evtgen}. 

For analyzing the simulated events it is important that the generated jet flavor is known~\cite{bid_nim}.
If a jet contains a simulated $b$ hadron, 
i.e. $\Delta R(\text{jet},\text{hadron})= \sqrt{(\Delta\phi)^2+(\Delta\eta)^2} <0.5$, it 
is flagged as a $b$ jet. If no $b$ hadron is contained within 
the jet, but a $c$ hadron is contained then it is defined as a $c$ jet. 
This sequence guards against cases where a $b$ quark decays to a $c$ quark. 
The remaining jets, which do not contain $b$ or $c$ hadrons, are defined as light jets.

\section{Tracking and primary vertex reconstruction}
\label{sec:reconstruction}

Past and current $b$ jet identification algorithms at \Dzero\ are based on three main inputs:
\begin{itemize}
\item Particle tracks: reconstructed from hits in the CFT and SMT tracking detectors
\item Vertices: reconstructed from at least two tracks originating from the same point
\item Calorimeter jets: reconstructed from their energy deposition in the calorimeter
\end{itemize}
After the track finding step we select the primary $p\bar{p}$ interaction vertex, from which we select
tracks for use in the identification algorithms (described in Sec.\thinspace\ref{sec:taggability}). 
These steps are briefly described below. A more detailed discussion of the various 
objects can be found in Ref.\thinspace\cite{bid_nim}.

\subsection{Track selection}                            
\label{sec:tracking}
For a track to be reconstructed it must first be detected with at least one hit in the SMT and 
at least six hits in the CFT for forward tracks and more than seven for central tracks. 
These tracks are also required to have transverse momentum $p_{T}^{\text{trk}}>0.5$~GeV and a 
distance of closest approach with respect to the the primary interaction vertex ($dca$) 
of less than 4 mm along the axis of the beam, $z$, and $2$ mm in the transverse plane with respect to the beam.

\subsection{Primary vertex reconstruction}
\label{sec:pv}

Knowledge of the $p\bar{p}$ interaction point is needed for the precise reconstruction 
and measurement of all objects in the calorimeter and provides an important point of reference 
for measuring lifetime based variables, which are discussed in Sec.\thinspace\ref{sec:variables}. 
Multiple interactions may occur during a single beam bunch crossing, 
making it necessary to identify the primary vertex (PV) associated with the interaction 
of interest. To form a PV candidate~\cite{bid_nim}: 
\begin{enumerate}[(i)]
\item two tracks must originate  less than 2 cm apart in the $z$ direction; 
\item an initial vertex fitting using a Kalman filter algorithm~\cite{kalman_filter} to obtain a list of candidate vertices; 
\item a second vertex fitting iteration using an adaptive algorithm to reduce the effect of outlier tracks;  
\item the PV is selected as the vertex with the lowest probability of originating from a soft underlying event.
\end{enumerate}
 
\subsection{Jet reconstruction and calibration}
\label{sec:jets}

Jets are reconstructed from energy deposits in the calorimeter using the iterative midpoint cone 
algorithm~\cite{jet_algo} with a cone of radius $R=0.5$. By design,
this algorithm provides reduced sensitivity to the presence of soft or collinear radiation from partons. 
The energies of jets are corrected for detector response, the presence of noise, multiple 
$p\bar{p}$~interactions, and for energy deposited outside of the jet reconstruction cone~\cite{jet_energy_correction}. 

\section{Algorithm prerequisites}
\label{sec:preliminaries}

Jets and their track information have to fulfill certain criteria, described below, before being used as inputs for $b$ jet identification. 

\subsection{Taggability}\label{sec:taggability}

Since $b$ jet identification algorithms are based solely on tracking and vertex information,
it is important to require that each jet reconstructed in the calorimeter is associated with tracks 
in the tracking system. We implement this ``taggability''~\cite{bid_nim} requirement separately from 
the requirements of the $b$ jet identification algorithm, allowing for the algorithm's performance to be less dependent on 
possible variations of the tracking system efficiency. 
For a jet reconstructed in the calorimeter to be considered {\it taggable} it must be matched to at least 
two tracks within a cone of radius $R=0.5$ with the origin set along the jet axis. 
All identification efficiencies and misidentification rates, which are the rates at which light jets are selected by the algorithm, 
are measured relative to {\it taggable} jets. 90\% the jets selected for this analysis with $p_{T} > 20$~GeV will be classified as taggable.  

\subsection{$V^{0}$ rejection}
\label{sec:v0}

Neutral hadrons containing strange quarks ($V^{0}$) have decay signatures 
similar to those of $b$ hadrons. In particular, $K_S$ and $\Lambda$ hadrons have 
lifetimes of $90$~ps and $263$~ps, respectively. 
To suppress this background, we reject secondary vertices with two oppositely charged tracks with the following criteria:

\begin{itemize}
\item The $z$ projection of each track must have a $dca < 1$~cm. This requirement suppresses mis-reconstructed tracks.

\item The significance of the $dca$, $S_{d}=dca/\sigma_{dca}$, of 
each track relative to the PV in the transverse plane has $|S_{d}|>3$.

\item The tracks associated with the $V^0$ candidate must have $dca < 200~\mu$m. 
This guarantees that $V^0$s from long lived neutral hadrons are 
rejected, not those which may have originated from $b$ hadron decays.

\item The invariant mass of the two tracks must be outside the mass range expected from 
$K_S$ or $\Lambda$, $472~\text{MeV}<m(\pi \pi)<516~\text{MeV}$ and 
$1108~\text{MeV}<m(p \pi)<1122~\text{MeV}$.
\end{itemize}

To reject photon conversions we reject pairs of tracks which have a negligibly small 
opening angle between an electron and positron in the plane transverse to the beam line. 
To be rejected the tracks from the electron and positron must be less than 30~$\mu$m apart at the
point where their trajectories are parallel to each other. In addition their invariant mass must be less than $25$~MeV.

\section{$\boldmath{b}$ jet identification algorithms}

For physics analyses prior to the year 2010 \dzero~used three algorithms based on charged tracks to identify $b$ jets~\cite{bid_nim}. 

\begin{description}
\item[{\bf Counting Signed Impact Parameters (CSIP)} -] 
CSIP determines the number of displaced tracks identified to a jet based on 
the $S_{d}$ of each track. To be selected by this algorithm a jet must have 
at least three tracks with $S_{d} > 2$, 
or two tracks with $S_{d} > 3$.

\item[{\bf Jet Lifetime Impact Parameter (JLIP)} -] 
The JLIP algorithm uses the IP of all tracks
associated with a jet to construct a probability that the jet is a
light flavor jet. The JLIP probability is constructed such that it
is uniformly distributed between 0 and 1 for light flavor jets,
while for heavy flavor jets the JLIP probability is close to zero.

\item[{\bf Secondary Vertex Tagger (SVT)} - ]
The SVT uses tracks that are significantly displaced from
the PV to reconstruct secondary vertices. A jet is tagged if it is
matched to a secondary vertex (SV),  $\Delta R(\text{jet, SV}) < 0.5$. This algorithm can
be tuned by varying the requirements on the tracks \pte,
$\chi^{2}$ per degree of freedom for the secondary vertex,
the transverse impact parameter significance of the tracks 
with respect to the primary vertex ($S_{xy}$),
and decay length significance of the secondary vertex in the plane transverse to the beam ($S_{dl}$).
These selections are optimized in a set of five SVT algorithms (SVT$1-5$)
that provide complementary information about the jet.
The track selections for the different configurations are listed in Table~\ref{tab:SVT}.
\end{description}

\begin{table}
 \caption{\label{tab:SVT} Track selection requirements for the five SVT algorithm configurations: 
 Super Loose (SVT1), Medium Loose (SVT2), Loose Extra (SVT3), Loose (SVT4), and Tight (SVT5).}
  \begin{center}\begin{tabular}{lcccccc}
      \hline
      \hline
       Track cuts & &SVT1 & SVT2 & SVT3 & SVT4 & SVT5 \\ \hline
       \pt [GeV]    & $>$ & 0.5 & 0.5 & 0.5 & 1 & 1 \\ 
       $\chi^{2}$ & $<$ & 15 & 15 & 10 & 10 & 3 \\ 
       $S_{xy}$  & $>$ & $-$ & 1.5 & 3 & 3 & 3.5 \\ 
       $S_{dl}$    & $>$ & $-$ & $-$ & 5 & 5 & 7 \\ \hline \hline
    \end{tabular}
   
    \label{tab:mlpparameters}
  \end{center}
\end{table}

In Ref.\thinspace\cite{bid_nim}, we described how input variables obtained from
these tools were combined using a neural network to construct the D0 NN-algorithm (D0-NN). 
The D0-NN shows significant performance improvements compared to the first-level algorithms.
In the following, we describe how further improvements have been achieved
using an extended set of input variables, making use of both decision trees and a neural network.
The new algorithm which results from these improvements is called \bl, standing for a multivariate
analysis that discriminates between $b$ quark and light jets.

\section{MVA$_{{\boldmath bl}}$ Algorithm}
\label{sec:BL}
To develop the \bl~algorithm we generate two MC samples: $10^{6}$ di-$b$ jet signal events 
and $10^{6}$ di-light jet background events. 
We use variables (discussed below) which separate $b$ jets from light jets to
train six random forests (RF) using the {\sc root tmva}~\cite{bib:tmva} framework. One RF is 
trained using the impact parameter properties from the CSIP and JLIP algorithms and one for each set of SVT variables extracted from 
the five different SVT algorithms configurations. 

These six RFs are then combined using a neural network implementation, 
the {\sc TMultiLayerPerceptron} (MLP), also within the {\sc root}~\cite{root} framework. This neural network 
utilizes the non-linear correlations between inputs to produce the \bl~output. 
This improves discrimination over the D0-NN by the inclusion of an order 
of magnitude more variables.

\subsection{Input variables}
\label{sec:variables}

\subsubsection{Impact Parameter Variables}
\label{sec:ip_variables}

To train the RF based on variables derived from the impact parameter properties we combine the following variables:  
\begin{enumerate}
\item the output of the JLIP algorithm; 
\item the output of the CSIP algorithm; 
\item the reduced JLIP~\cite{bid_nim}, which is computed by removing the track with the 
lowest probability of originating from the PV and then recalculating the JLIP; 
\item the combined probability~\cite{bid_nim} associated with the tracks with the highest 
and second highest probability of coming from the PV; 
\item the largest separation in $\Delta R=\sqrt{\Delta\phi^2+\Delta\eta^2}$ 
between any two tracks within a jet, $\max[\Delta R(\mbox{tracks})]$; 
\item the sum of the $\Delta R$ distances between each track matched to the jet and the center of the calorimeter jet, 
$\Sigma_{\text{trk}}\Delta R(\text{trk, jet})$; 
\item the $p_{T}$-weighted $\Delta R$ width of the tracks relative to the calorimeter jet defined as
\begin{equation}
\Theta \equiv \frac{\sum\limits_{\text{trk}}p_{T}^{\text{trk}}\times\Delta R(\text{trk, jet})}{\sum\limits_{\text{trk}} p_{T}^{\text{trk}}};
\end{equation}
\item the total transverse momentum of all tracks in the jet cone; 
\item the total number of tracks matched to the jet. 
\end{enumerate}
The resulting RF output distribution is displayed in Fig.\thinspace\ref{fig:RFoutput}(a).

\subsubsection{Secondary Vertex Variables}
\label{sec:svt_variables}

The SVT algorithms preselect a set of tracks according
to their kinematic properties and reconstruction quality. 
As a consequence, starting from a common set of tracks, the various SVT configurations
lead to different secondary vertices with different properties providing a complementary
set of variables for each jet.
We then train five RFs using variables associated with the secondary vertices.

\begin{figure*}
\includegraphics[width=0.48\textwidth]{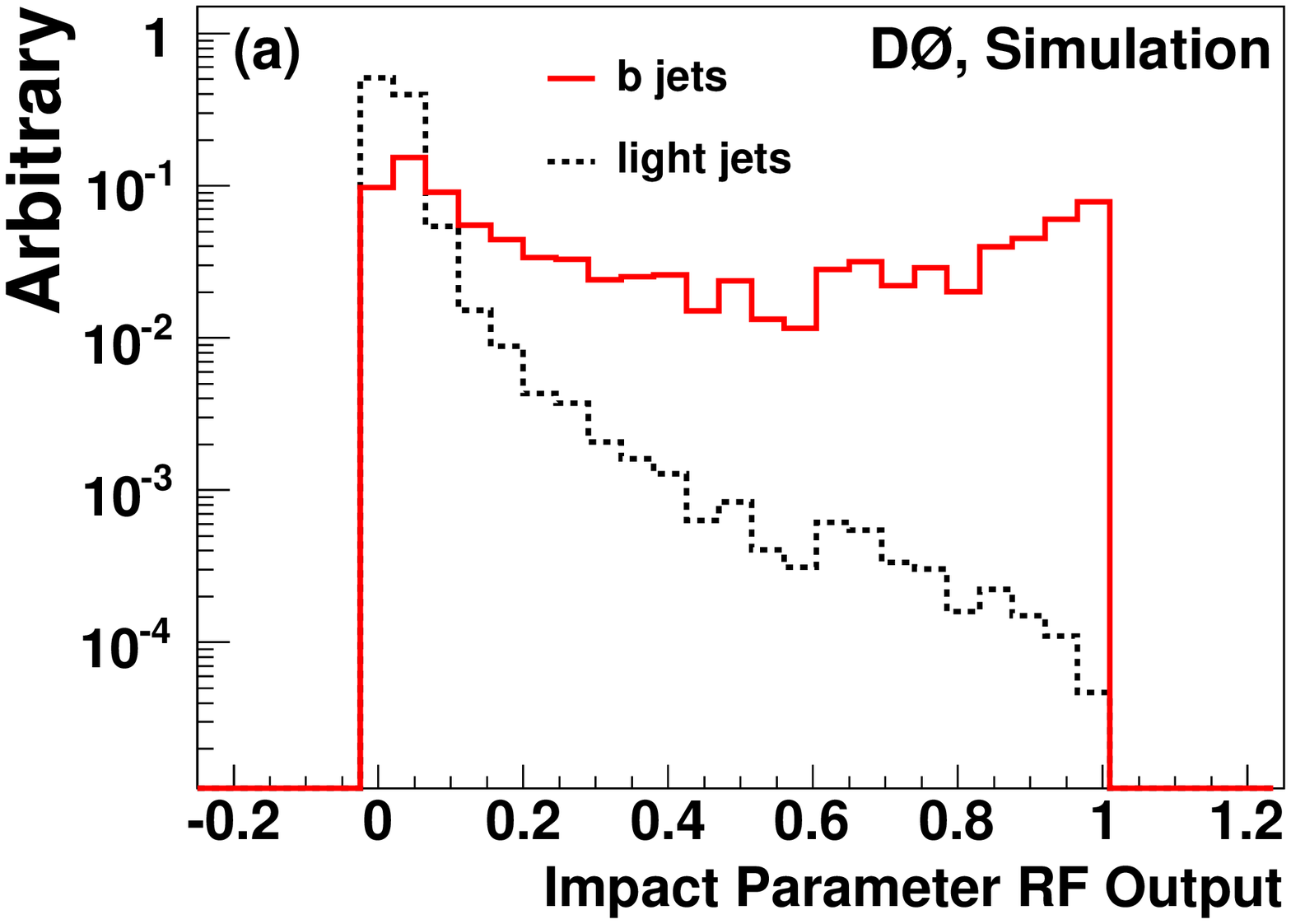}\includegraphics[width=0.48\textwidth]{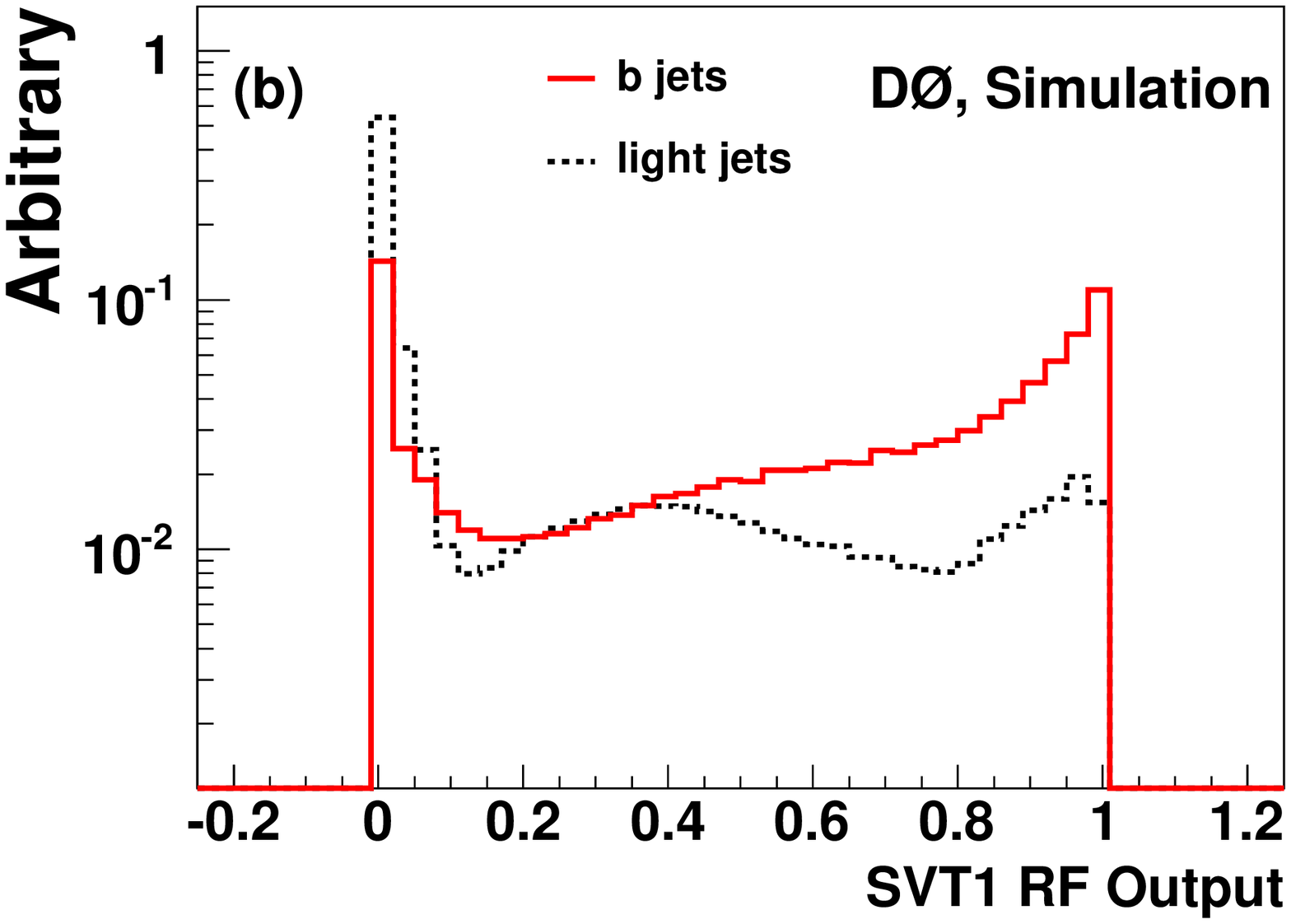}\\
\includegraphics[width=0.48\textwidth]{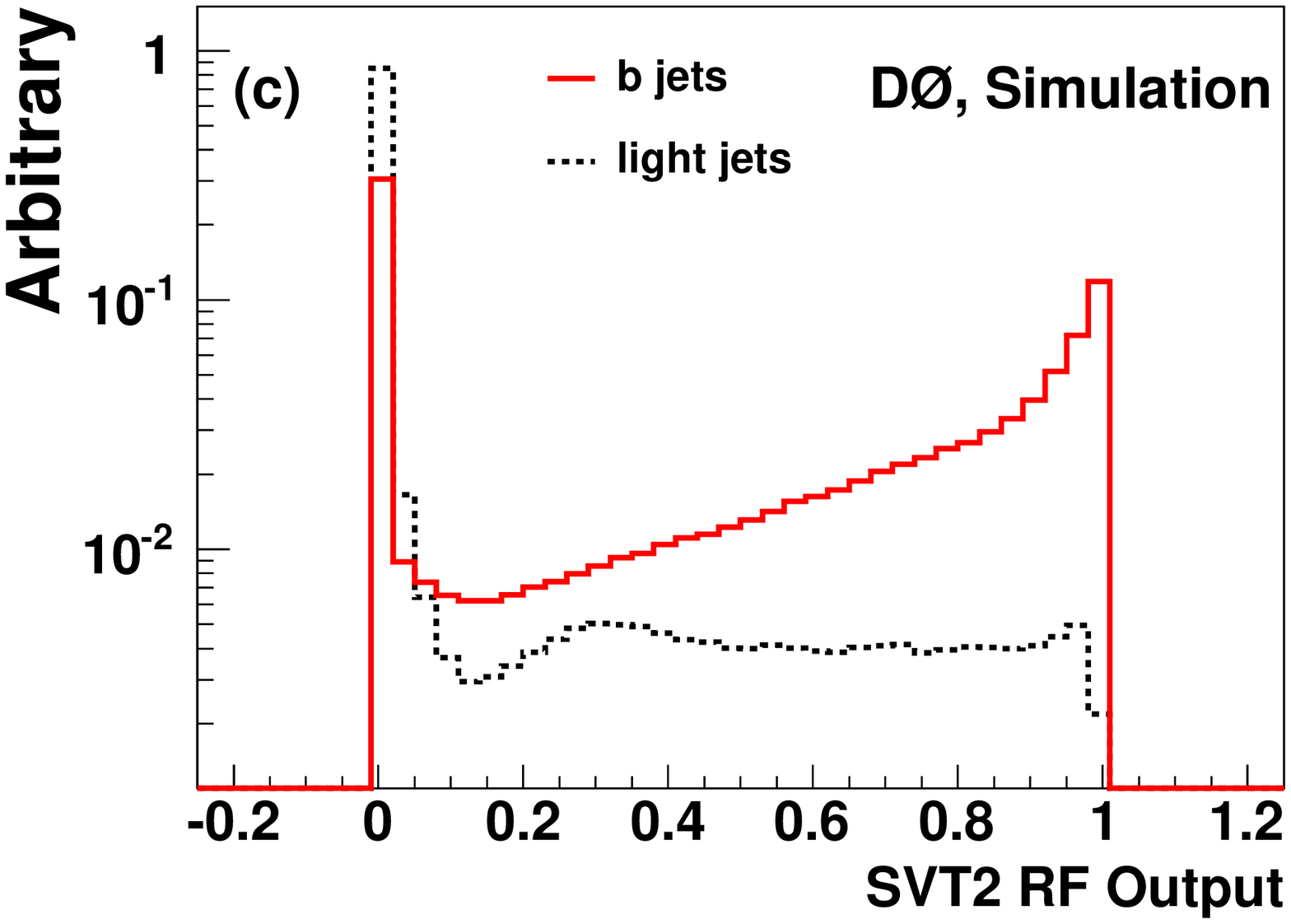}\includegraphics[width=0.48\textwidth]{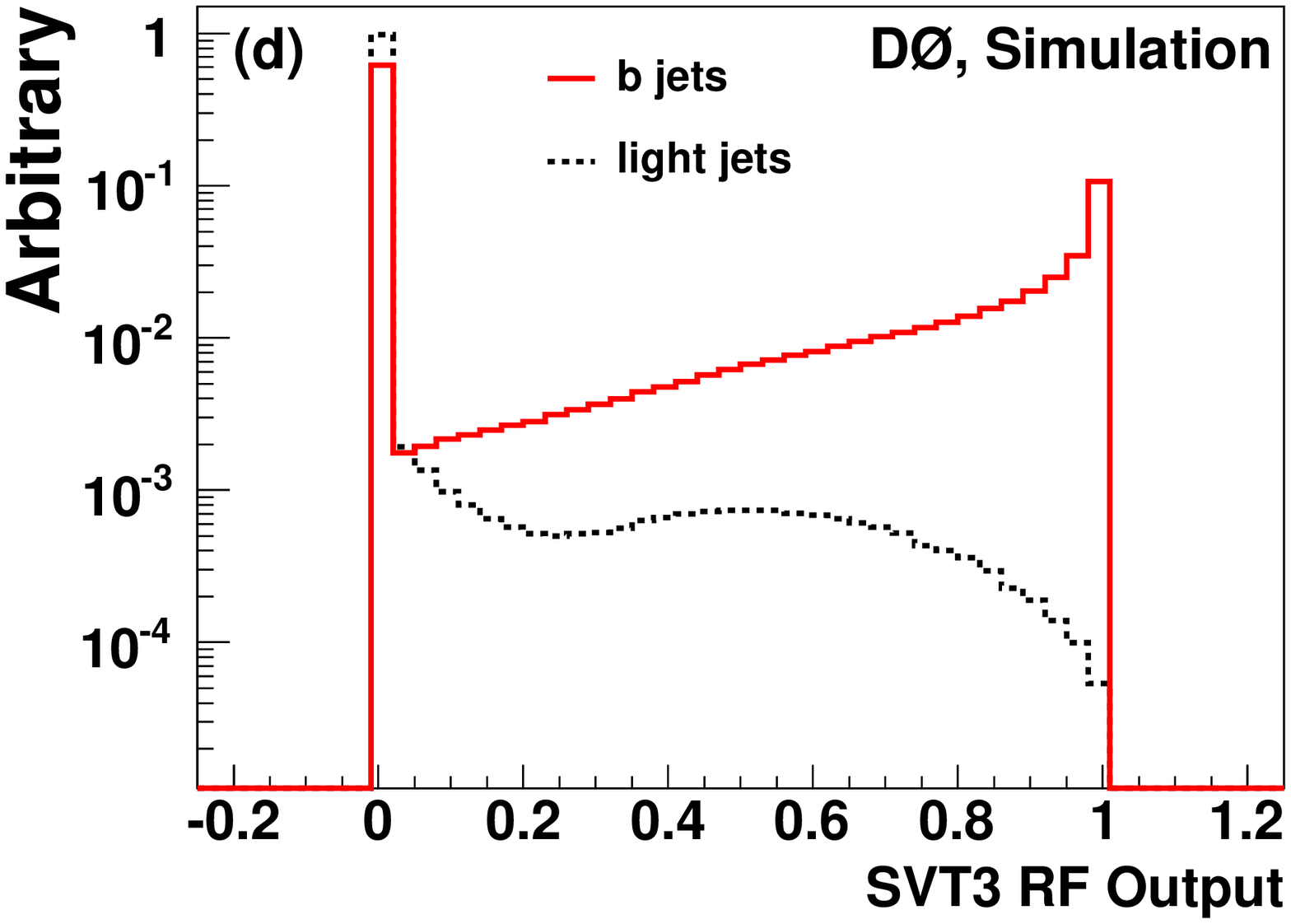}\\
\includegraphics[width=0.48\textwidth]{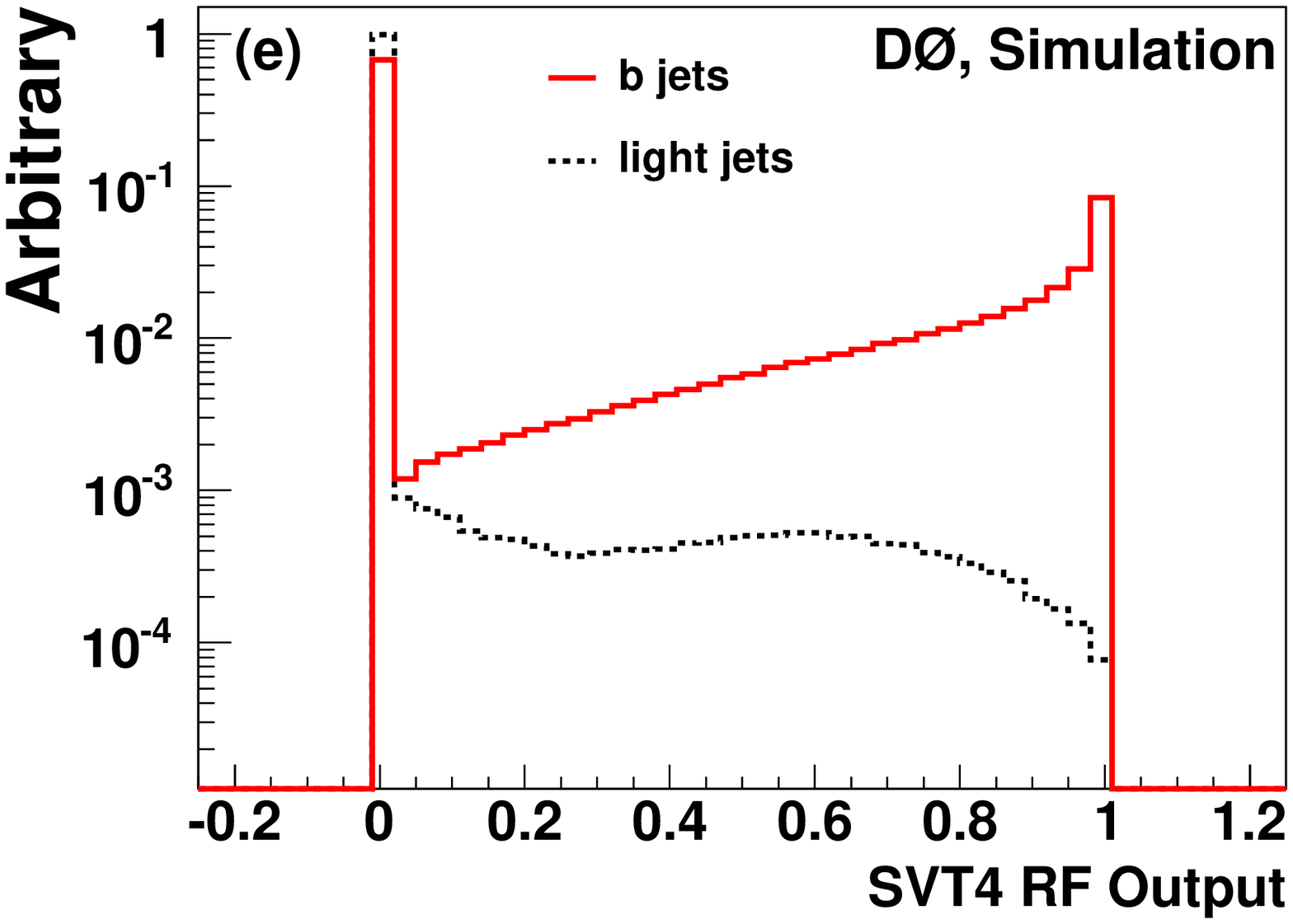}\includegraphics[width=0.48\textwidth]{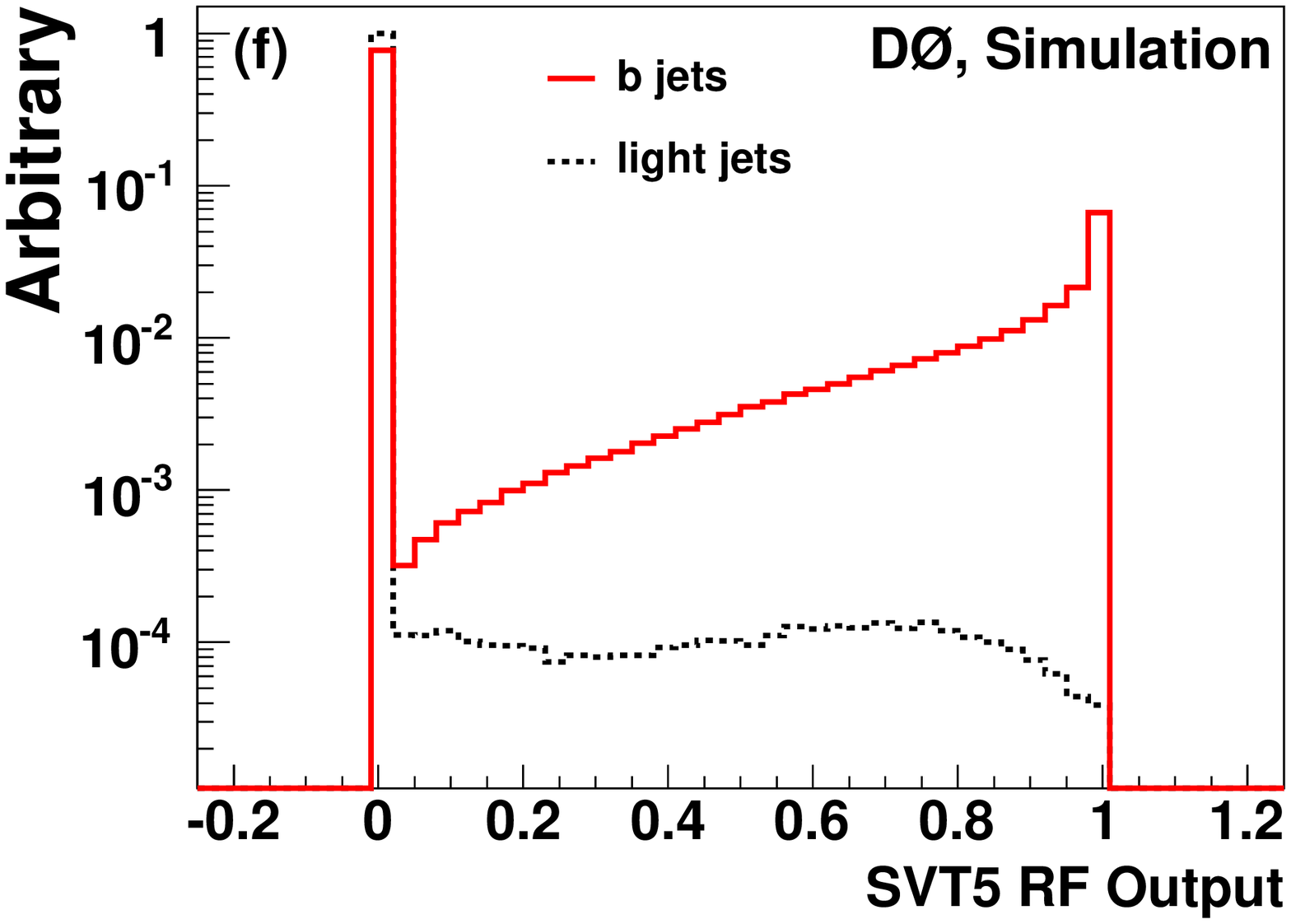}\\
\caption{(color online) Distributions of the six RF outputs for (a) the impact parameter variables 
and (b$-$f) the five configurations of the SVT algorithm.} 
\label{fig:RFoutput} 
\end{figure*}

In total each of the SVT RFs uses 29 input variables: 
\begin{enumerate}
\item the $p_{T}$ of the highest $p_{T}$ track matched to the secondary vertex, $p_{T}^{1}$; 
\item the $p_{T}$ of the second highest $p_{T}$ track matched to the secondary vertex, $p_{T}^{2}$;
\item the $p_{T}$ fraction carried by the tracks from the secondary vertex tracks, $p_{T}^{\text{SVT}}/p_{T}^{\text{jet}}$; 
\item the number of tracks originating from the secondary vertex; 
\item the mass of the secondary vertex (\msv), calculated by summing 
all track four-momentum vectors assuming that all tracks originate from pions; 
\item the signed decay length significance of the secondary vertex in the plane transverse to the beam direction; 
\item the JLIP probability of the tracks matched to the secondary vertex; 
\item the sum of $\chi^{2}/\mbox{n.d.f.}$ of the tracks matched to the secondary vertex; 
\item the number of secondary vertices which can be reconstructed from the tracks matched to the jet; 
\item the signed IP of the track with the highest momentum measured transverse to the direction of the secondary vertex;
\item the number of tracks matched to the jets; 
\item The proper lifetime of the secondary vertex, computed using \msv, in the plane transverse to the beam direction; 
\item the decay length of the secondary vertex in the plane transverse to the beam direction; 
\item the decay length of the secondary vertex in the beam direction; 
\item the $p_{T}$ of the highest $p_{T}$ track in the jet divided by the $p_{T}$ of the secondary vertex ($p_{T}^{\text{SVT}}$), $p_{T}^{1}/p_{T}^{\text{SVT}}$;
\item the $p_{T}$ of the second highest $p_{T}$ track normalized 
to the secondary vertex $p_{T}$, $p_{T}^{2}/p_{T}^{\text{SVT}}$; 
\item the $dca$ of the secondary vertex to the PV in the plane transverse to the beam; 
\item the $dca$ of the secondary vertex to the PV in the beam direction; 
\item the $p_{T}$ of the track which has the highest momentum measured relative 
to the direction of the secondary vertex; 
\item the momentum of the secondary vertex in the plane transverse to the calorimeter jet direction; 
\item the $p_{T}$ of the highest $p_{T}$ track divided by the total jet $p_{T}$, $p_{T}^{1}/p_{T}^{\text{jet}}$; 
\item the $p_{T}$ of the second highest $p_{T}$ track divided by to total jet $p_{T}$, $p_{T}^{2}/p_{T}^{\text{jet}}$; 
\item the angle between the tracks emerging from the secondary vertex projected into the plane transverse to the beam direction;
\item the angle between the tracks emerging from the secondary vertex projected in the beam direction; 
\item the $\Theta$ (as defined above) as measured for tracks matched to the secondary vertex; 
\item the $\max[\Delta R(\mbox{tracks})]$ of the tracks matched to the secondary vertex; 
\item the $p_{T}$ weighted charge ($q$) of the jet, measured as $\sum\limits_{\text{trk}}p_{T}^{\text{trk}}q^{\text{trk}}/p_{T}^{\text{jet}}$;
\item the signed decay length significance of the secondary vertex in the beam direction;
\item the radius of the cone enclosing all the tracks matched to the secondary vertex.
\end{enumerate}
The outputs of the five SVT RFs are shown in Figs.\thinspace\ref{fig:RFoutput}(b$-$f). 

\subsection{Optimized \bl~parameters}

The outputs of the six RFs, shown in Fig.\thinspace\ref{fig:RFoutput}, are combined
using an MLP neural network into a single variable. The training parameters
for the six separate RFs and the final MLP are optimized to minimize
the misidentification rate for a fixed $b$ jet identification efficiency. 
The RF parameters are the number of trees in the forest~(5) and
the number of variables considered at each random split~(all). 
The parameters used for building the final neural network discriminant
are the number of nodes~(7 input, 1 hidden, and 1 output) and the
number of training iterations~(50).

\subsection{ \bl~performance in simulation}
\label{sec:OPs}

\begin{figure*}\centering
\includegraphics[width=0.48\textwidth]{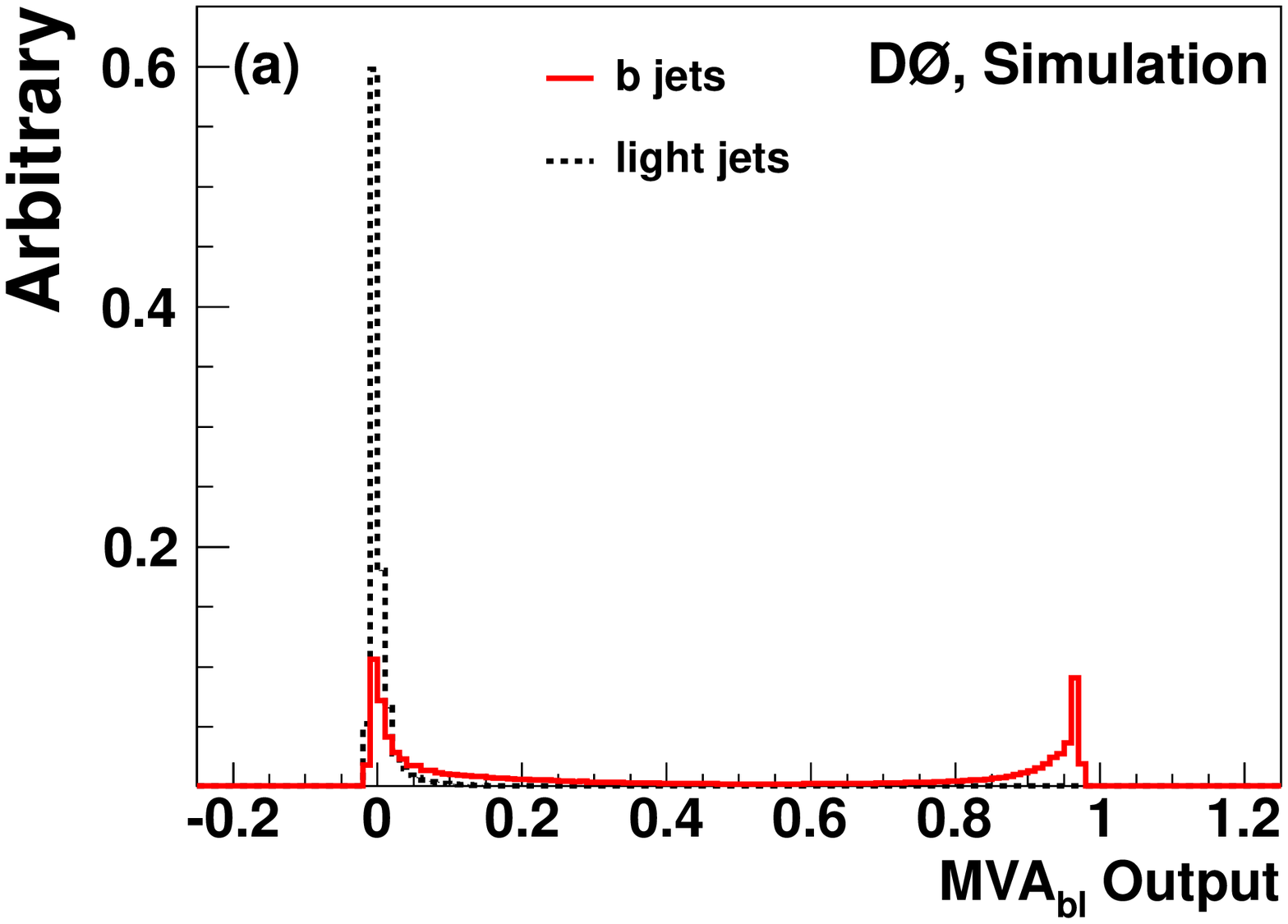}\includegraphics[width=0.48\textwidth]{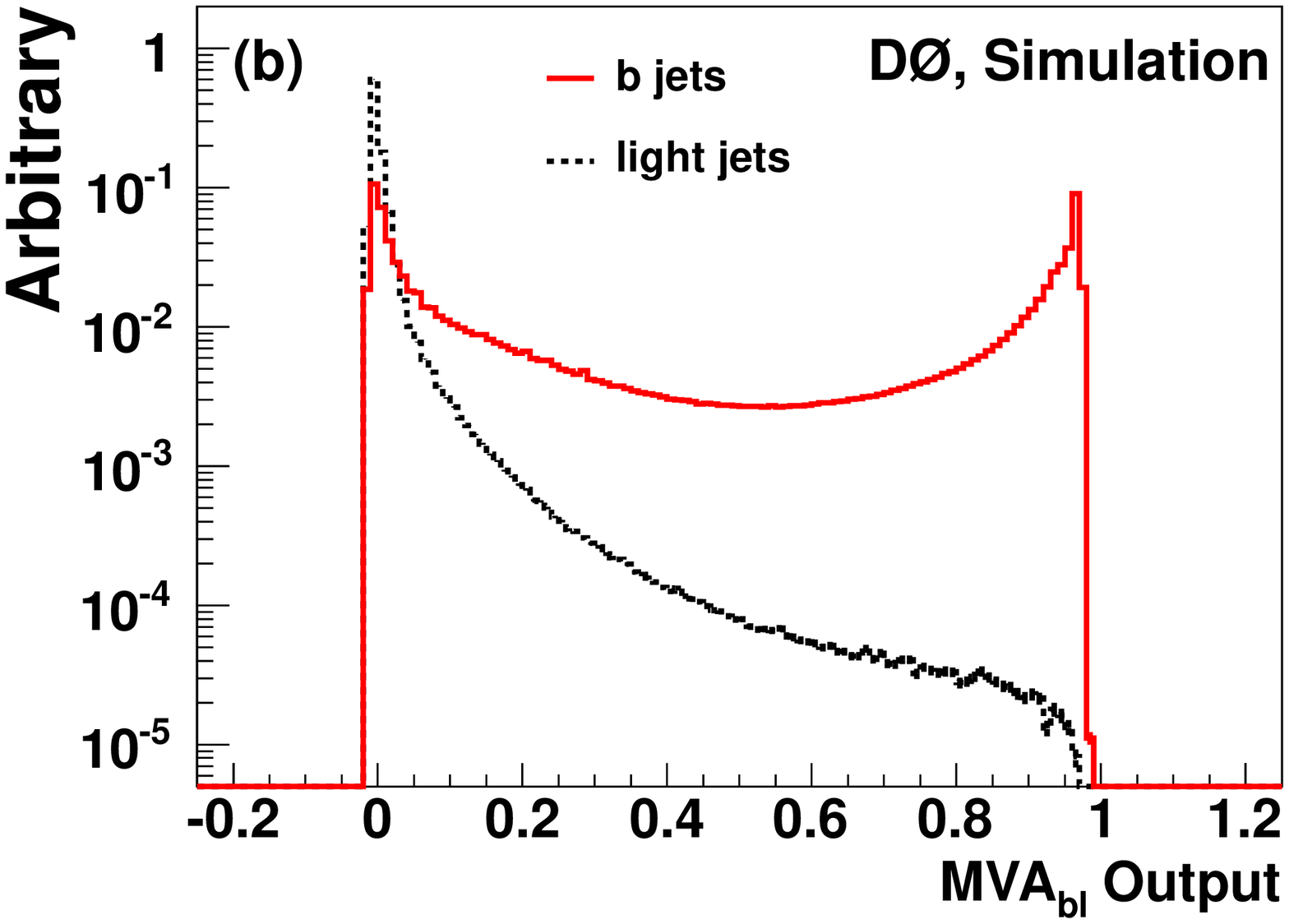}
\caption{(color online) The \bl~output for light flavored and $b$ jets in MC events, with (a) linear and (b) logarithmic scales. Both distributions are normalized to unity.} \label{fig:mvabl}
\end{figure*}

\begin{figure}\centering
\includegraphics[width=0.48\textwidth]{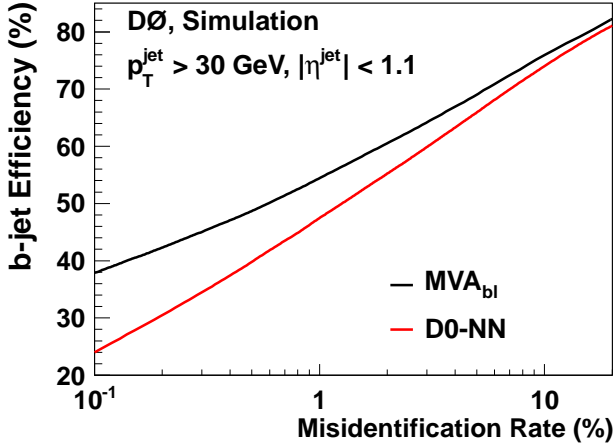}
\caption{(color online) The performance of the \bl~and D0-NN algorithms for jets with $|\eta^{\text{jet}}|<1.1$ 
and $p_T^{\text{jet}}>30$ GeV.} \label{fig:perf}
\end{figure}

\begin{figure*}\centering
\includegraphics[width=0.48\textwidth]{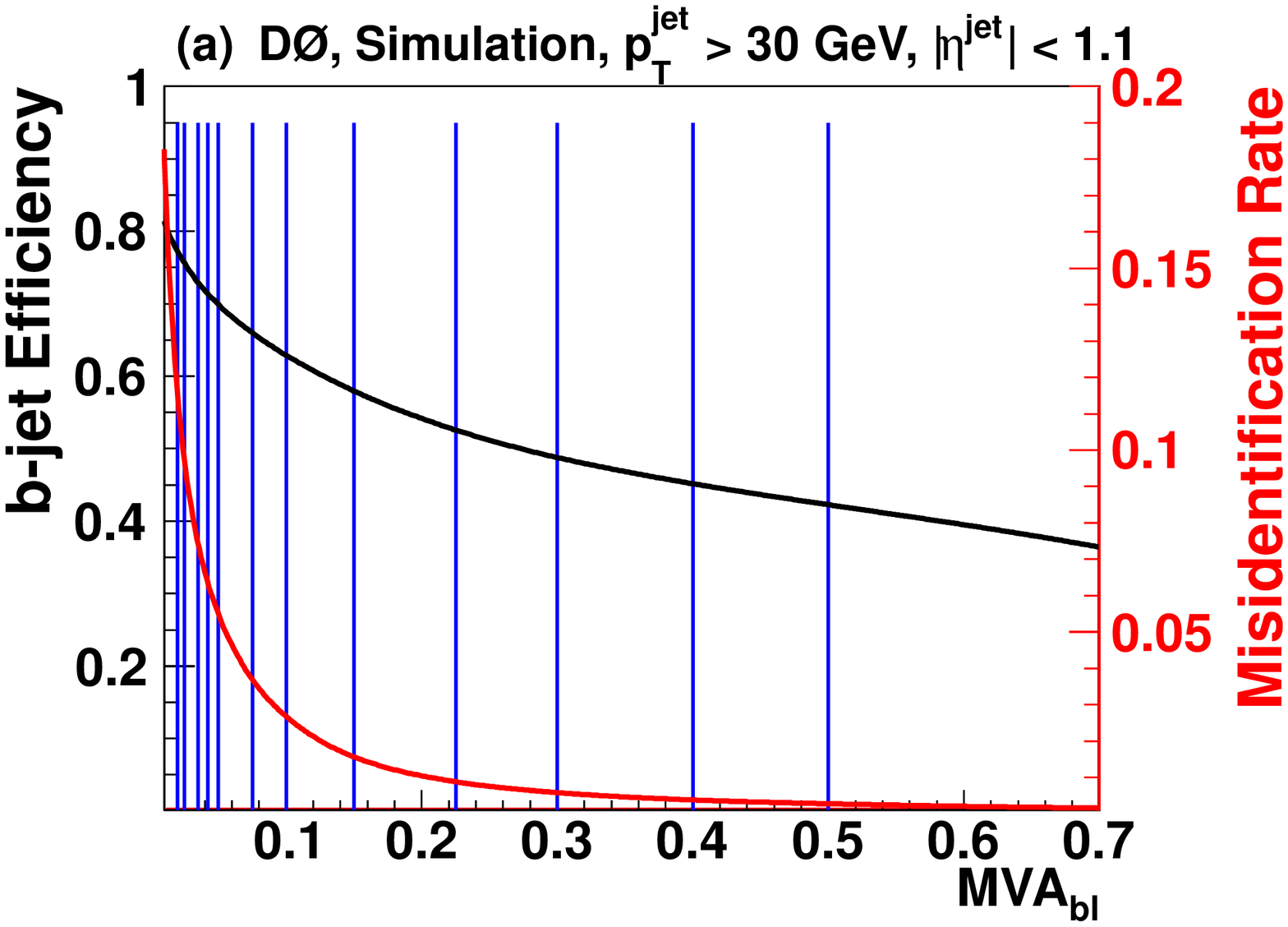}~~\includegraphics[width=0.48\textwidth]{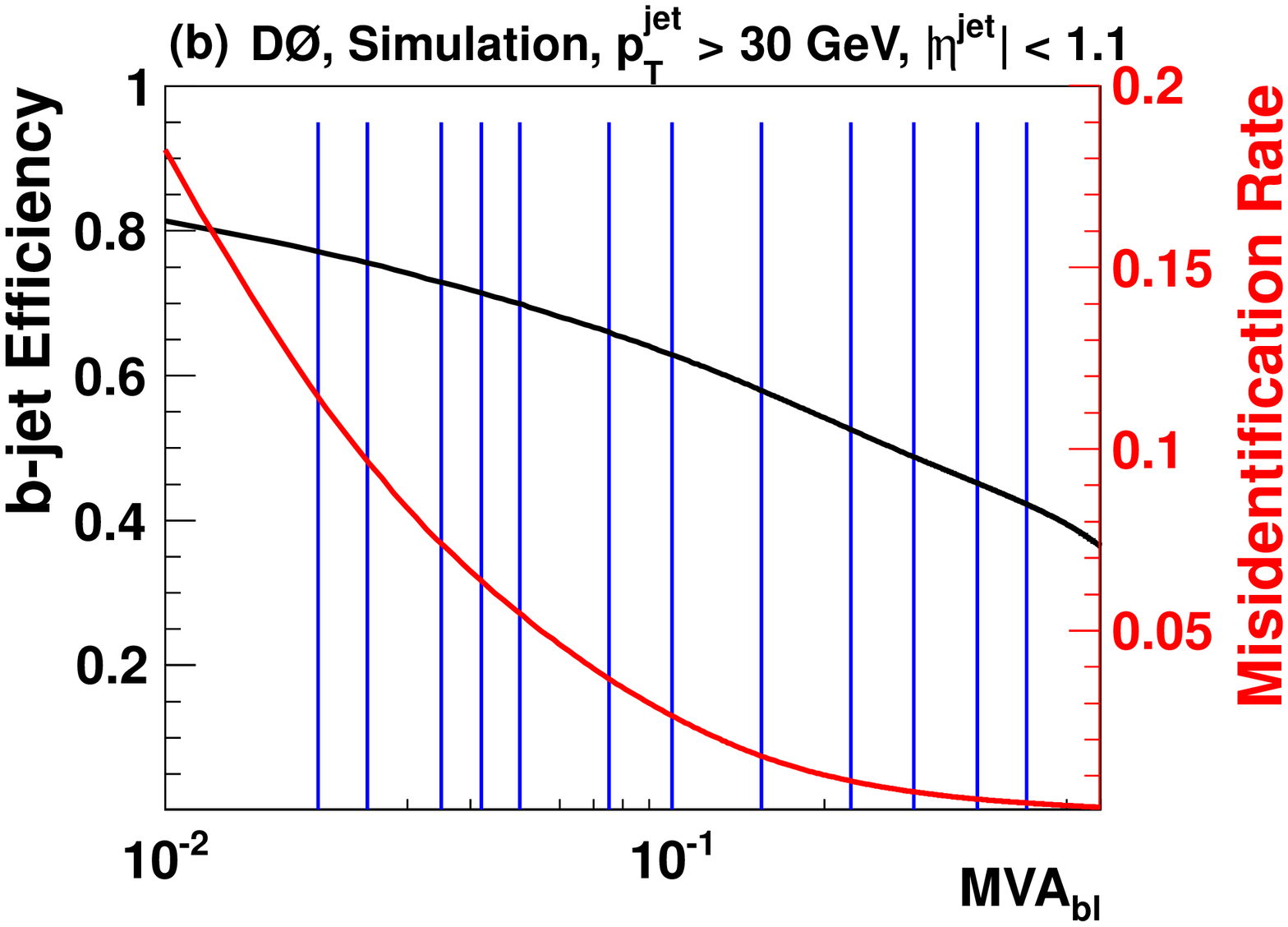}
\caption{(color online) The efficiency for selecting a $b$ jet and the light jet misidentification rate 
as a function of the \bl~requirement as determined in simulations. The vertical lines 
correspond to the selected operating points described in Sec.\thinspace\ref{sec:OPs}, with (a) linear and (b) logarithmic scales.} \label{fig:effvbl}
\end{figure*}

The performance of the \bl~algorithm is presented in Fig.\thinspace\ref{fig:mvabl}.
A measure of the discriminating power is given by the performance profile, or the identification 
efficiency of a $b$ jet versus the misidentification rate.
The comparison of the  performance of the D0-NN and \bl~algorithms is presented in Fig.\thinspace\ref{fig:perf}.
At low values of the misidentification rate, the \bl~preforms significantly better than the D0-NN, 
while at high values they are similar. 
We define a set of benchmark points, designated as operating points (OPs) below, and determine the efficiency and 
misidentification rates of the OPs for use in subsequent analyses. For the \bl~algorithm, these points are 
defined in the following way:
\begin{description} \centering
\item L6, \bl~$> 0.02$; 
\item L5, \bl~$> 0.025$;
\item L4, \bl~$> 0.035$;
\item L3, \bl~$> 0.042$;
\item L2, \bl~$> 0.05$;
\item Loose, \bl~$> 0.075$; 
\item oldLoose, \bl~$> 0.1$; 
\item Medium, \bl~$> 0.15$;
\item Tight, \bl~$> 0.225$;
\item VeryTight, \bl~$> 0.3$;
\item UltraTight, \bl~$> 0.4$; 
\item MegaTight, \bl~$> 0.5$.
\end{description}
These OPs are displayed in Fig.\thinspace\ref{fig:effvbl} where the 
identification efficiency for $b$ jets and the misidentification 
rate for light jets are shown as a function of the \bl~output 
for simulated events.

\section{Efficiency Estimation}
\label{sec:efficiency}

Once the algorithm has been defined and its performance is quantified in simulation,
we compare the performance measured in data. 
This is a two step-process where we use the efficiencies 
in both data and MC to correct the simulation. 

\subsection{System8 method}

Using the \syst~(S8) formalism, the $b$ jet identification efficiencies 
can be measured directly from data~\cite{bid_nim}. 
A system of eight equations with eight unknowns is
constructed so that solution to these nonlinear equations
includes the efficiency for selecting $b$ jets. 

To determine the efficiency of identifying a $b$ jet we construct
a heavy flavor enriched data sample. These events contain two
back-to-back jets satisfying $|\Delta \phi(\text{jet}_1,\text{jet}_2)|>2.5$, one jet must 
have $p_{T} > 15$~GeV and $|\eta| < 2.5$ and 
be matched to a muon inside a cone of $R= 0.5$ around its centroid (called a muonic jet). 
The matched muon must have $p_{T}^{\mu}>4$~GeV. 
These events, now enriched in heavy flavor
jets, contain contamination from light jets due to muonic decays of $\pi^{\pm}$ and $K^{\pm}$.
Since the S8 method only accommodates a single background we combine the 
$c$ and light jet backgrounds into a single sample referred to as ``$cl$ jets".

Three additional requirements, or ``tags", are individually applied 
to muonic jets to create subsamples that are further enriched in $b$ jets. 
The first {\it tag} selects muonic jet that passes a given \bl~OP 
(described in Sec.\thinspace\ref{sec:OPs}). 
The second {\it tag} is a requirement on $p_{T}^{\mu}$
relative to the direction obtained by adding the muon and jet momenta, known as $p_{T}^{rel}$. 
Requiring that $p_{T}^{rel}>0.5$~GeV removes light jets as the large 
$b$ quark mass leads to large muon $p_{T}^{rel}$~\cite{Hedin}.
The final {\it tag} is a requirement that the jet which is recoiling from the muonic jet has $\text{JLIP} < 0.005$, this is known as the ``away-side tag". 
The ``away-side tag" allows us to select a data sample heavily enriched in pair-produced back-to-back $b$ jets.
Using the JLIP to {\it tag} this {\it away} jet leads to an enrichment in the overall heavy flavor content
without applying any additional requirements on the muonic jet.
The following coefficients are introduced into the S8 formulation to
account for possible correlations between these {\it tags}:

\begin{enumerate} [I]
  \item[$\beta$]: Correlations between the {\it away tag} and \bl~requirements for $b$ jets.
  
  \item[$\alpha$]: Correlations between the {\it away tag} and \bl~requirements for $cl$ jets.
  
  \item[$\kappa_{b}$]: Correlations between the $p_{T}^{rel}$ and \bl~requirements for $b$ jets.
  
  \item[$\kappa_{cl}$]: Correlations between the  $p_{T}^{rel}$ and \bl~requirements for $cl$ jets.
  \end{enumerate}

The above {\it tags} are denoted as $k$, for the \bl~requirement; $m$, for the $p_{T}^{rel}$ requirement; and, $b$, for the {\it away} tag. These are applied both individually and concurrently and will appear as superscripts in the following system of S8 equations:

\begin{eqnarray}
  \begin{array}{lclcl}
    f_{b}&+&f_{cl} &=&1 \\
    f_{b} \varepsilon_{b}^{k} &+& f_{cl} \varepsilon_{cl}^{k}  &=& Q^{k} \\
    f_{b} \varepsilon_{b}^{m} &+& f_{cl} \varepsilon_{cl}^{m}  &=& Q^{m}\\
    f_{b} \varepsilon_{b}^{n} &+& f_{cl} \varepsilon_{cl}^{n}  &=& Q^{n}\\
    f_{b} \kappa_{b}\varepsilon_{b}^{k} \varepsilon_{b}^{m} &+& f_{cl} \kappa_{cl} \varepsilon_{cl}^{k}\varepsilon_{cl}^{m} &=& Q^{k,m}\\
    f_{b} \varepsilon_{b}^{m}\varepsilon_{b}^{n} &+&f_{cl} \varepsilon_{cl}^{m}\varepsilon_{cl}^{n} &=& Q^{m,n}\\
    f_{b} \beta\varepsilon_{b}^{n}\varepsilon_{b}^{k} &+&f_{cl} \alpha\varepsilon_{cl}^{n}\varepsilon_{cl}^{k} &=& Q^{n,k}\\
    f_{b} \kappa_{b}\beta\varepsilon_{b}^{k} \varepsilon_{b}^{m} \varepsilon_{b}^{n} &+& f_{cl}
    \kappa_{cl}\alpha\varepsilon_{cl}^{k} \varepsilon_{cl}^{m}\varepsilon_{cl}^{n} &=& Q^{k,m,n},
  \end{array}
\end{eqnarray}

\noindent where the subscripts $b$ and $cl$ refer either to $b$ or $cl$ jets,
$Q$ refers to the fraction of the total number of selected jets in the sample
that pass a given {\it tag}, $f_{X}$ denotes the fraction of events of
a given flavor $X$ in the initial un-tagged sample, and $\varepsilon^{Y}_{X}$ 
refers to the efficiency of a jet of flavor $X$ passing {\it tag} $Y$. $Q$ is determined 
from the data and $\alpha$, $\beta$, $\kappa_{b}$, and $\kappa_{cl}$ are determined
from simulations~\cite{bid_nim}. 
This leaves eight remaining unknowns which form the solution, including the
variable we are interested in: $\varepsilon^{k}_{b}$, the efficiency of a $b$ jet passing the \bl~requirement.
These equations give two possible solutions for $\varepsilon_{b}^{Y}$
but this can be resolved by requiring that 
$\varepsilon_{b}^{Y} > \varepsilon_{cl}^{Y}$.

The $b$ jet identification efficiency obtained with the S8 method is valid for 
muonic jets. To obtain the efficiency for inclusive $b$ jet decays, 
a correction factor is determined by using two samples of
simulated $b$ jets: muonic and inclusive. 
The final efficiency is then defined as
\begin{equation}
\varepsilon^{\text{data}}_{b} = \frac{\varepsilon^{\text{data}}_{b\rightarrow \mu X}}{\varepsilon^{MC}_{b\rightarrow \mu X}} \times \varepsilon^{MC}_{b} = SF \times \varepsilon^{MC}_{b}
\end{equation}
where $SF =  \varepsilon^{\text{data}}_{b\rightarrow \mu X}/\varepsilon^{MC}_{b\rightarrow \mu X}$ 
is the data-to-simulation efficiency correction factor, 
$\varepsilon^{\text{data}}_{b\rightarrow \mu X}$ is the efficiency for passing all \bl~OPs as measured by the S8,
and $\varepsilon^{\text{MC}}_{b\rightarrow \mu X}$ is the efficiency measured in simulation. 
The identification efficiency for $c$ jets is not measured directly from the data. 
It is assumed that the data-to-simulation scale factor is
identical for $b$ and $c$ jets~\cite{bid_nim}. The $c$ jet identification efficiency
is then derived from the simulation as 
\begin{equation}
\varepsilon^{\text{data}}_{c} = SF \times \varepsilon^{MC}_{c}.
\end{equation}

\subsection{\bl~efficiency}

Using this methodology we are able to determine $\varepsilon^{data}_{b}$ for the set of OP requirements.
We have selected two OPs, Loose and Tight, for demonstration.

\begin{figure*}\centering
\includegraphics[width=0.45\textwidth]{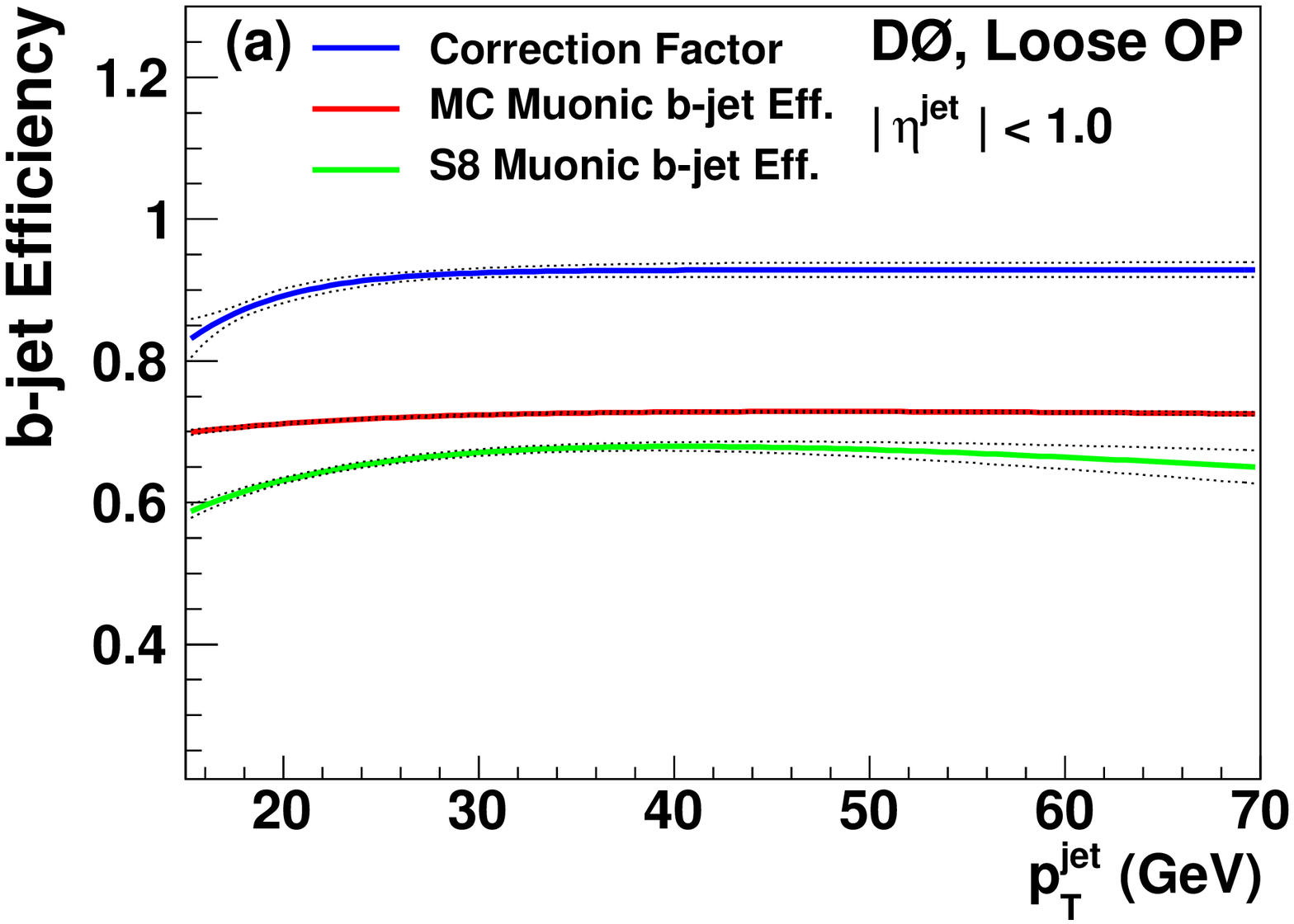} 
\includegraphics[width=0.45\textwidth]{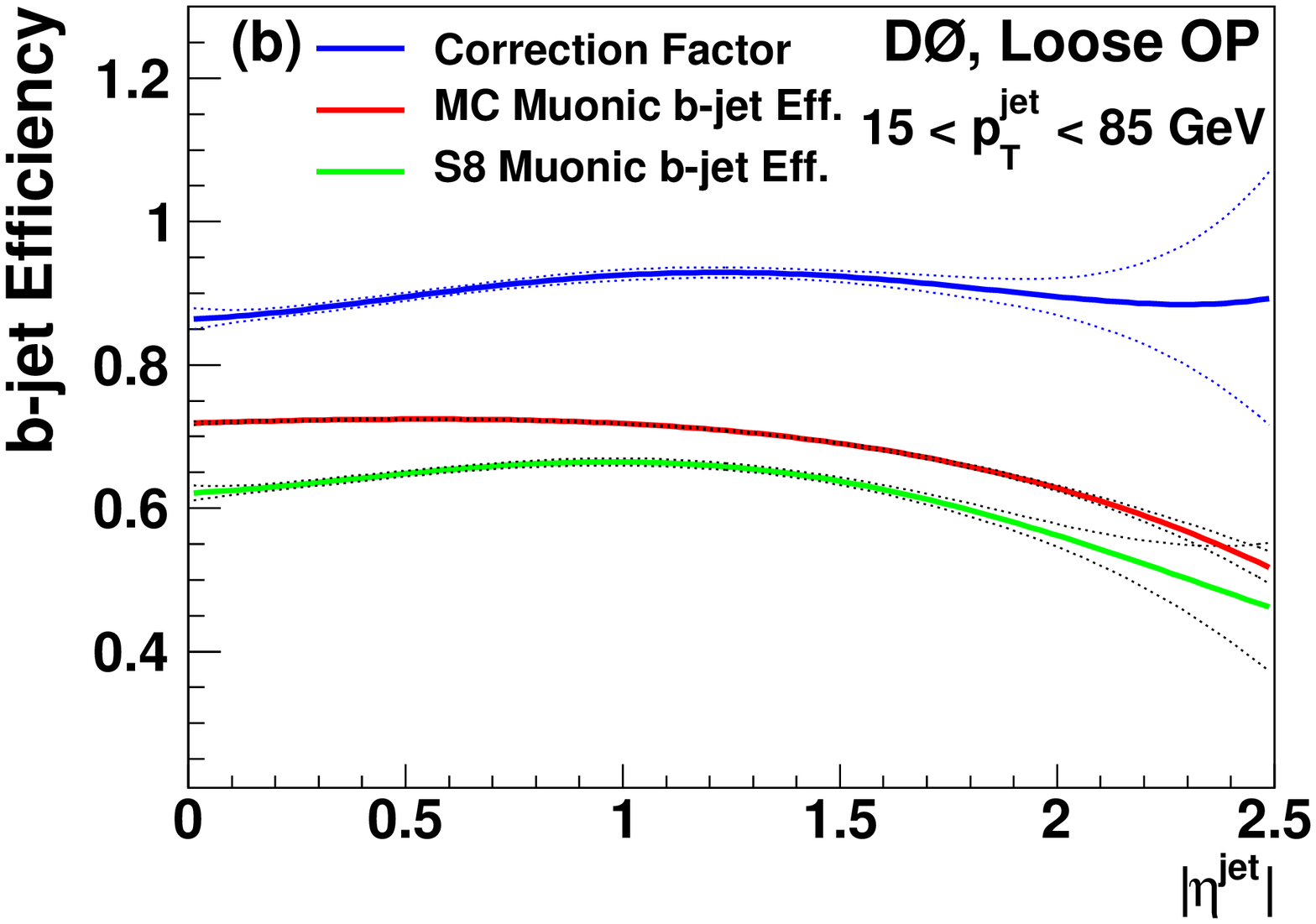}\\
\includegraphics[width=0.45\textwidth]{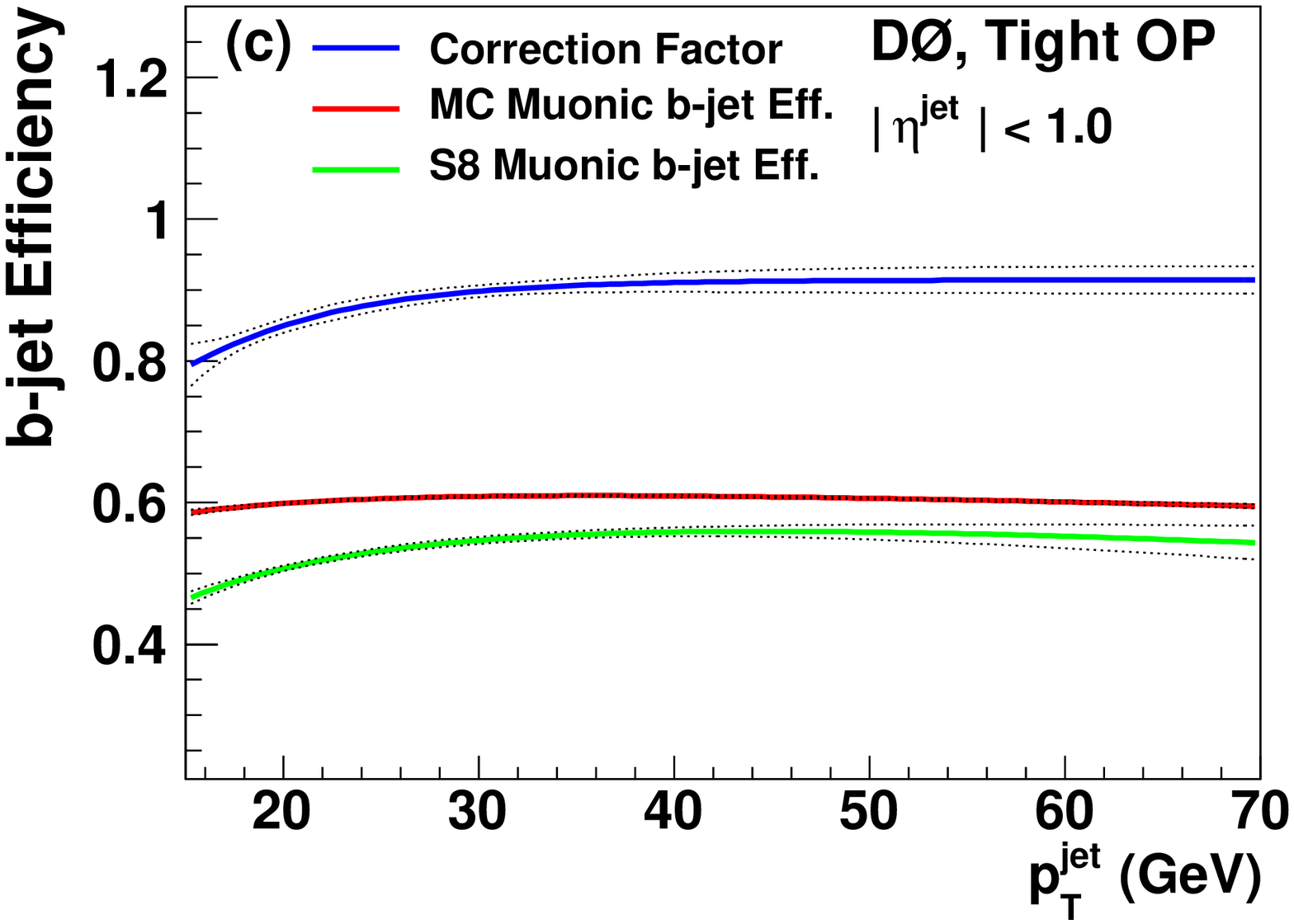} 
\includegraphics[width=0.45\textwidth]{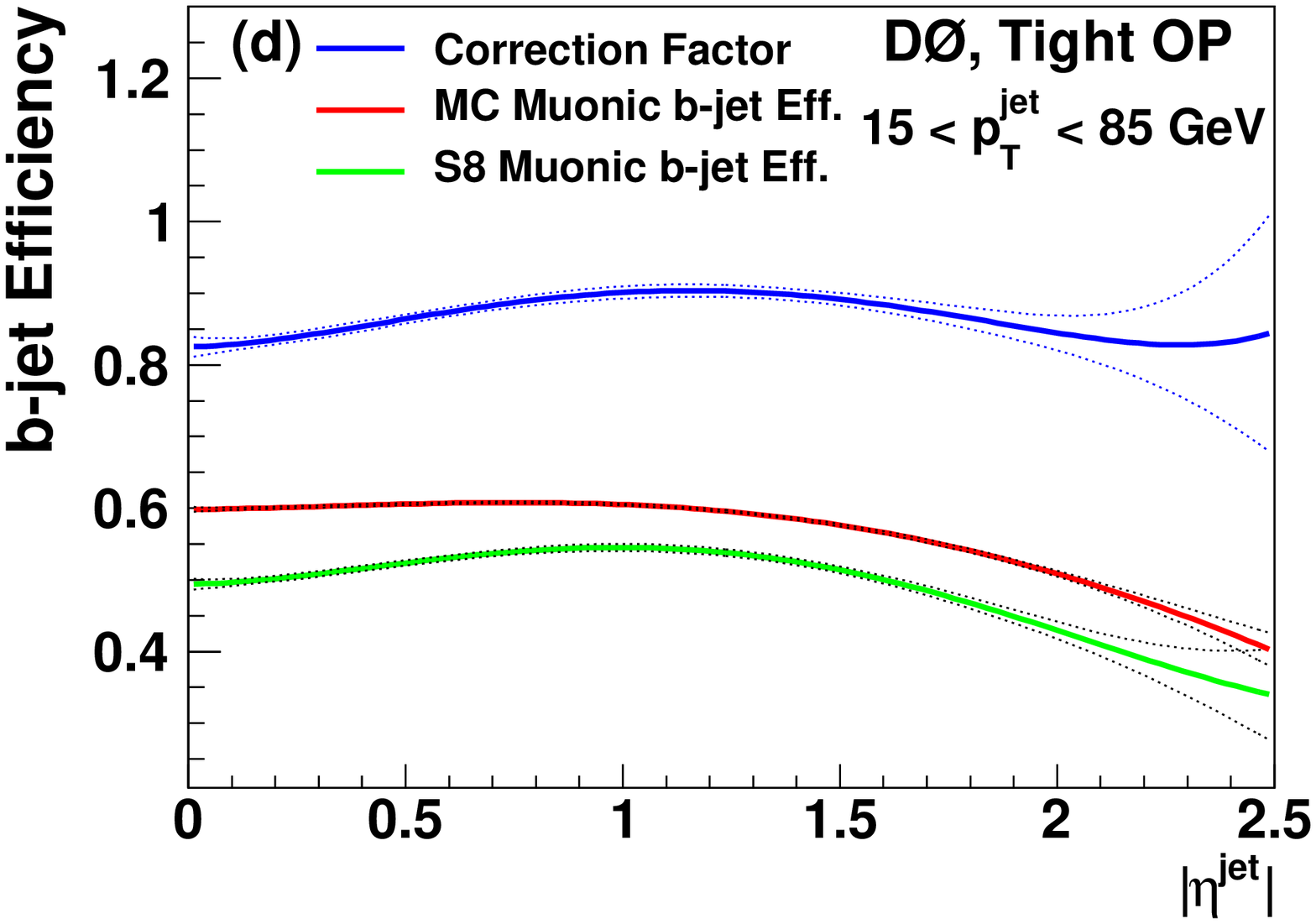}
\caption{(color online) The efficiency for selecting a muonic $b$-jet in MC and data using the S8 method. The correction factor, $SF$, which is used to model the algorithm's efficiency, is also shown. Two OPs are shown (a,b) the Loose and (c,d) Tight. The efficiencies are parameterized as a function of (a,c) $p_T$, for central jets and versus (b,d) $\eta$. The band which surrounds the lines corresponds to $\pm1\sigma$ total uncertainties.} \label{fig:hSF}
\end{figure*}

\begin{figure*}\centering
\includegraphics[width=0.45\textwidth]{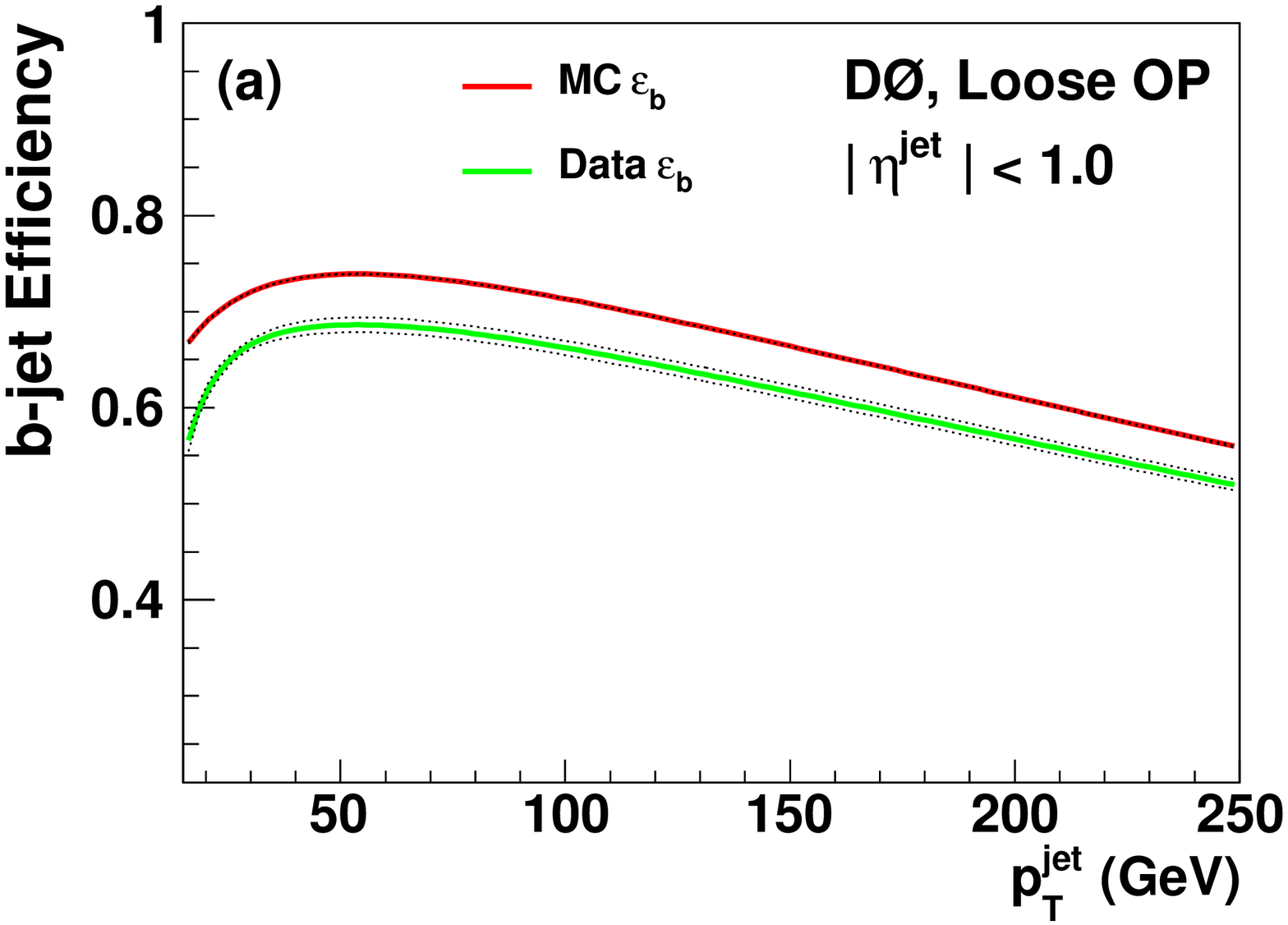} 
\includegraphics[width=0.45\textwidth]{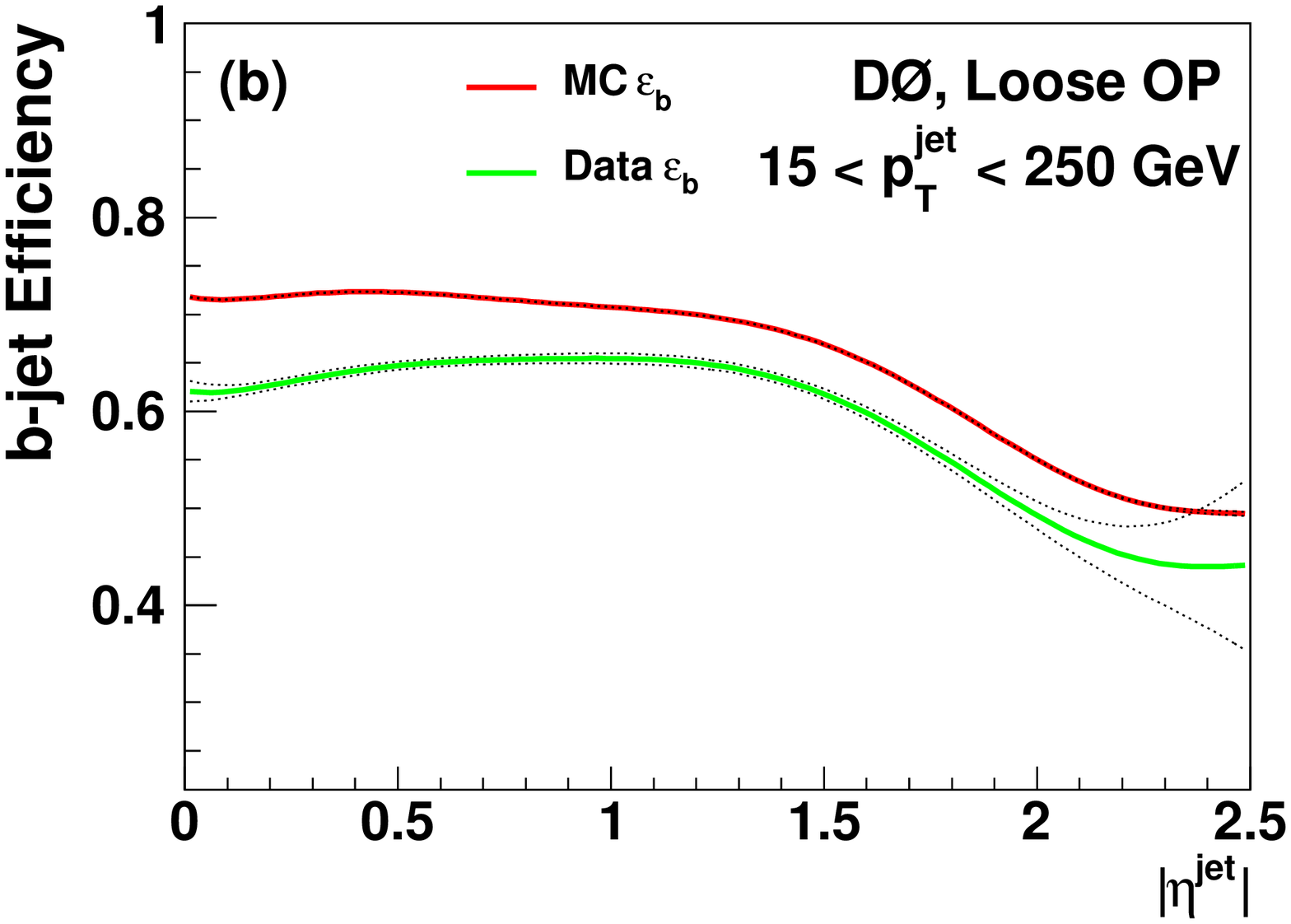}\\
\includegraphics[width=0.45\textwidth]{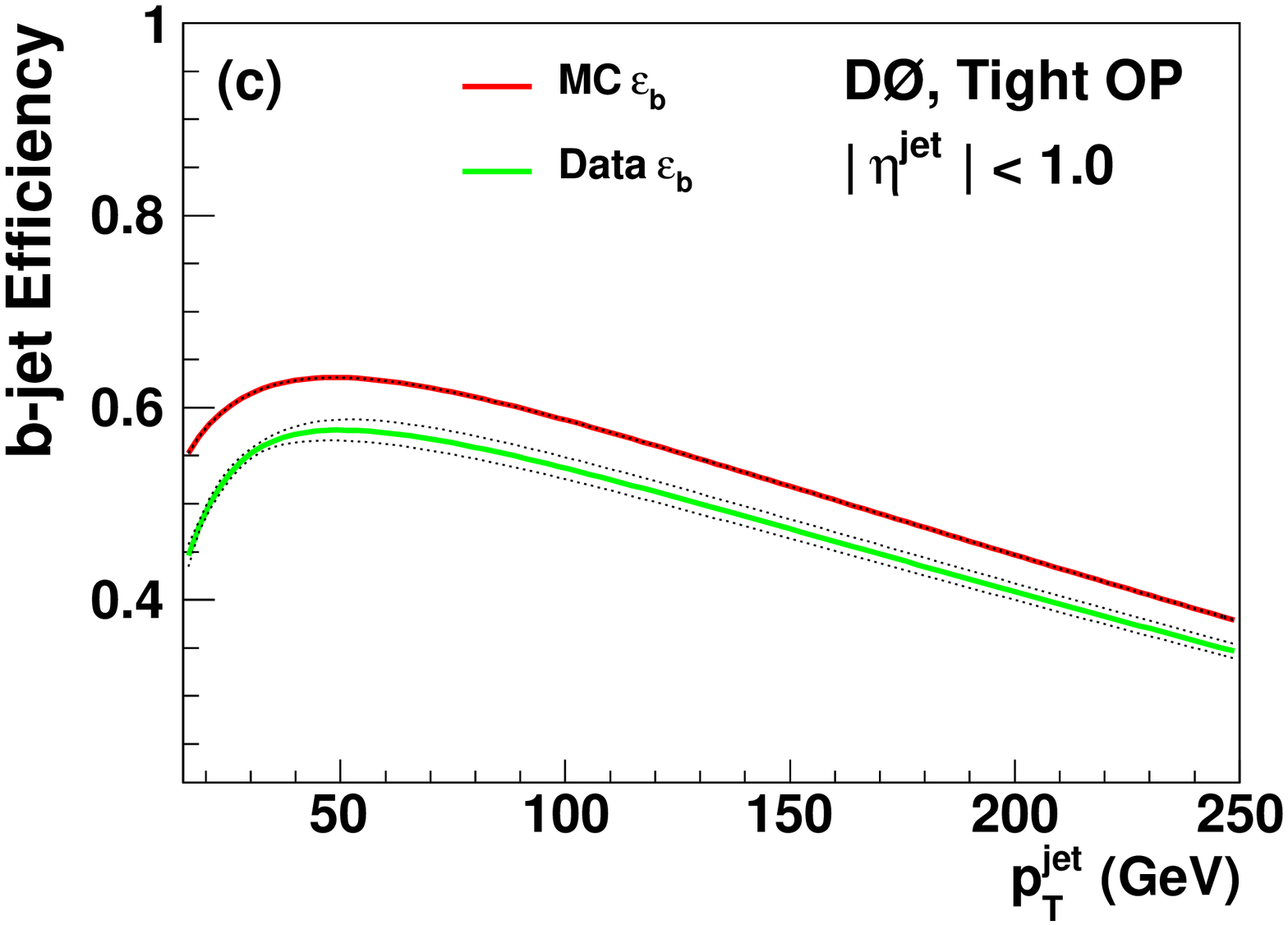} 
\includegraphics[width=0.45\textwidth]{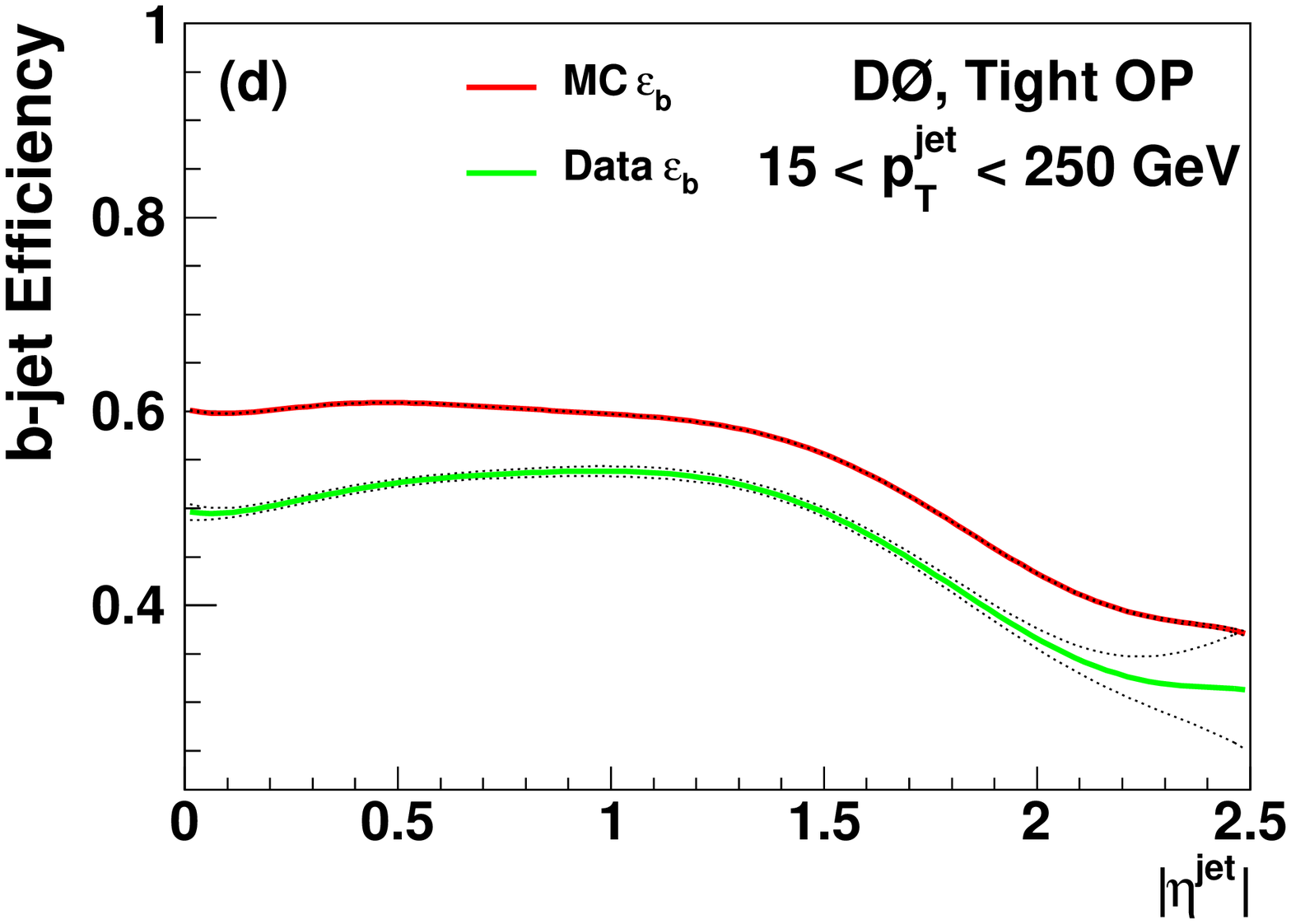}
\caption{(color online) The MC $b$ jet identification efficiency, as measured in dijet events along with the data $b$ jet identification efficiency. Two OPs are shown (a,b) the Loose and (c,d) Tight. The efficiencies are parameterized as a function of (a,c) $p_T$, for central jets and versus (b,d) $\eta$.} \label{fig:bTRFs}
\end{figure*}

\begin{figure*}\centering
\includegraphics[width=0.48\textwidth]{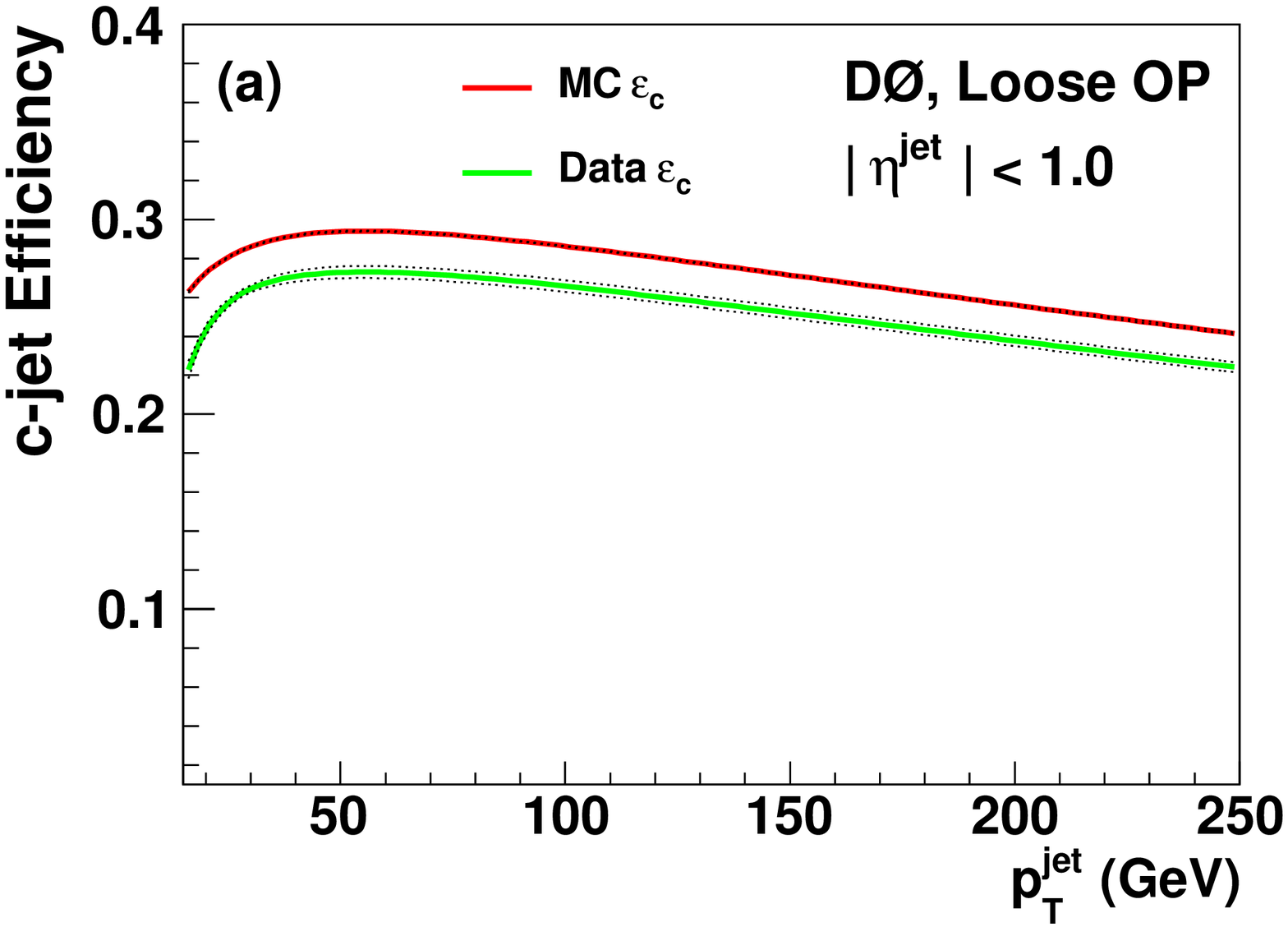} 
\includegraphics[width=0.48\textwidth]{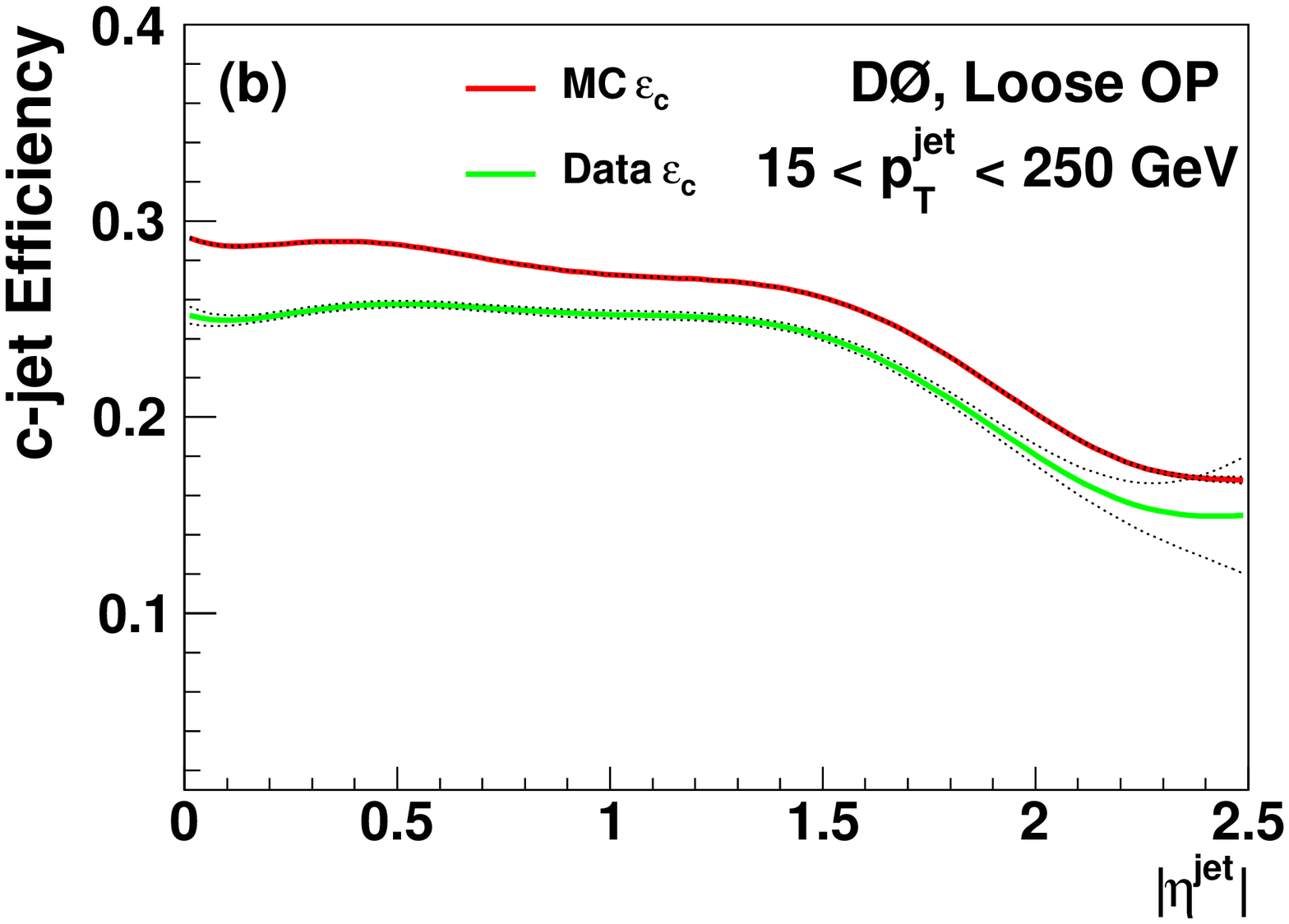}\\
\includegraphics[width=0.48\textwidth]{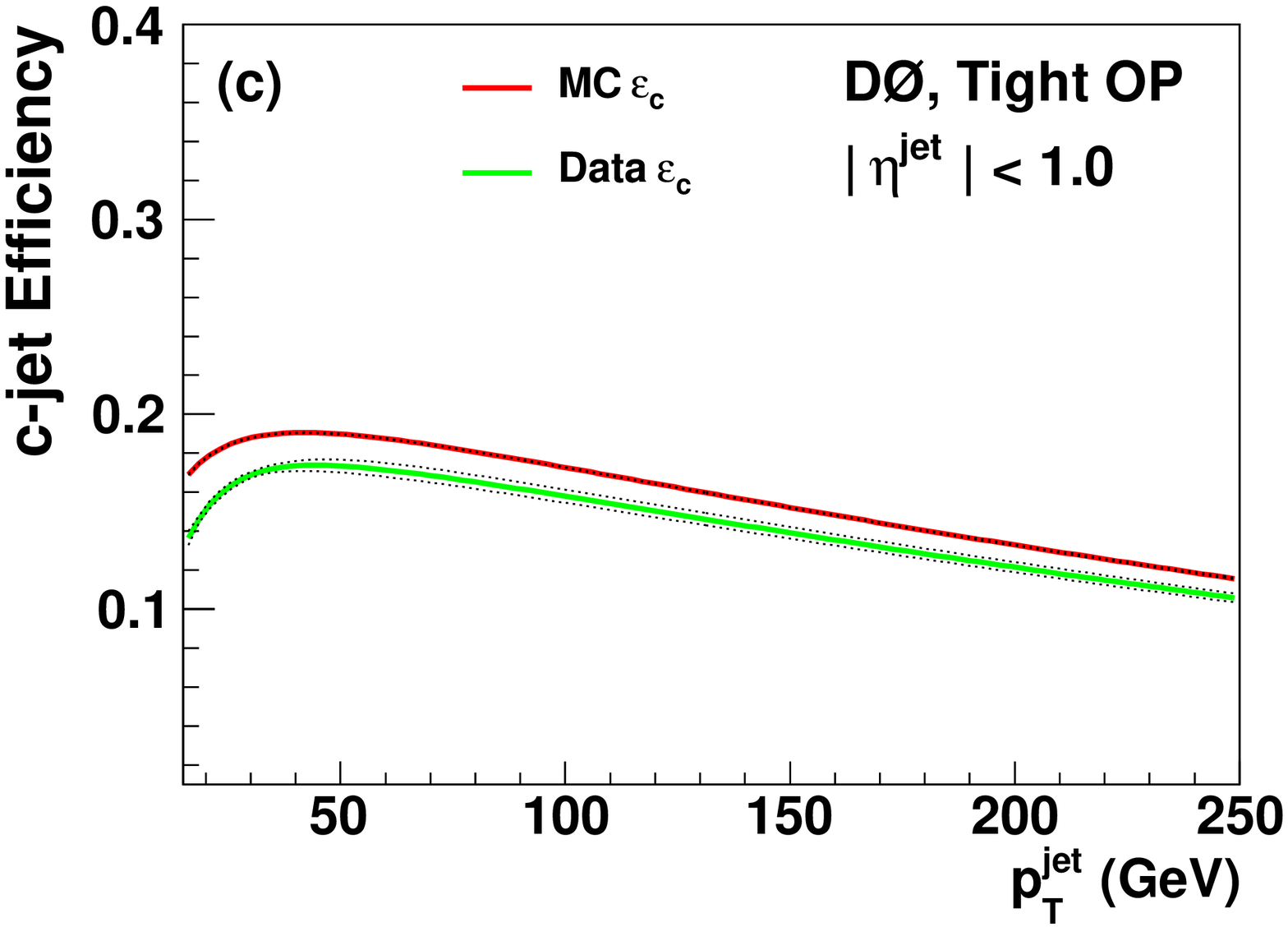} 
\includegraphics[width=0.48\textwidth]{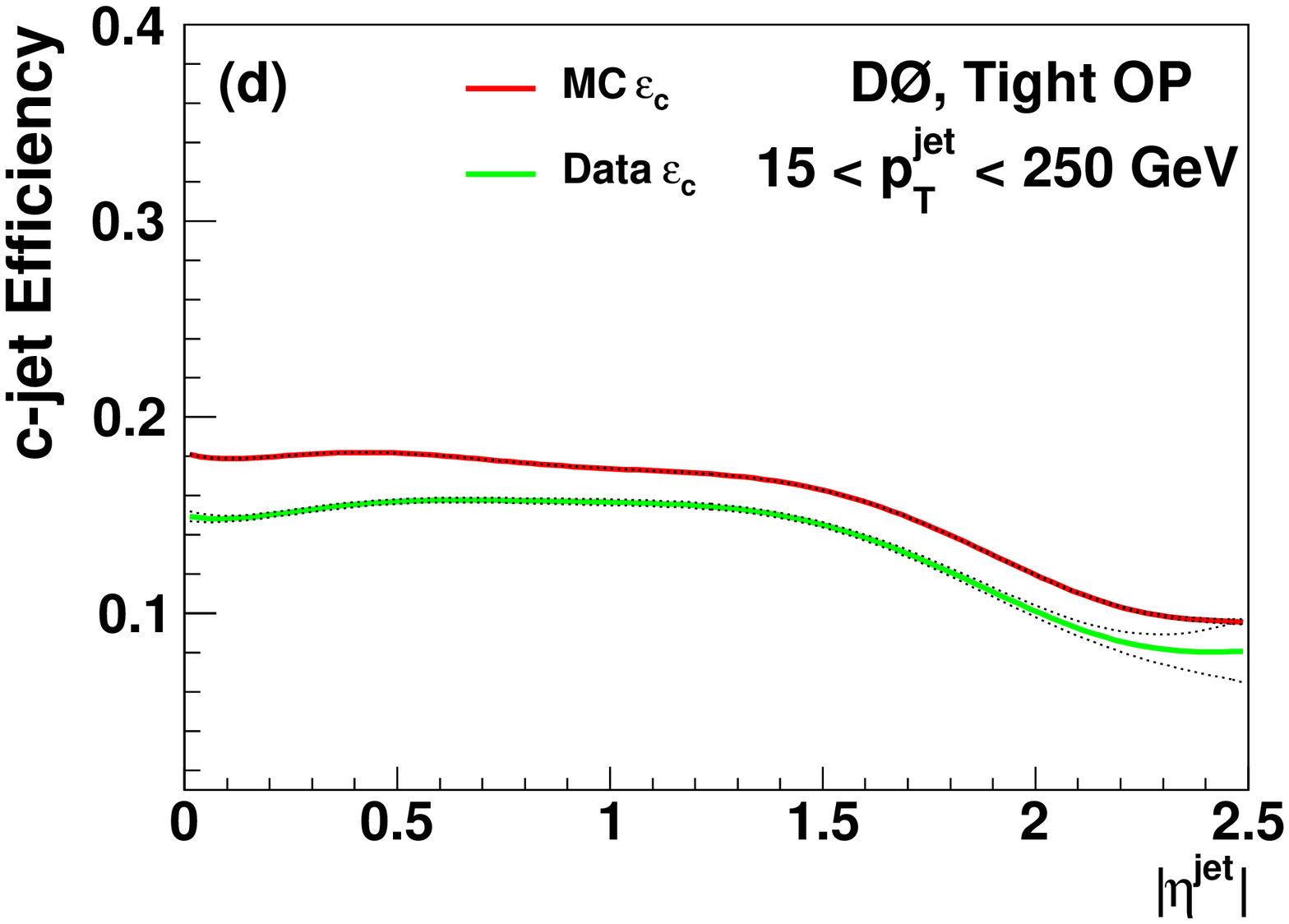}
\caption{(color online) The MC $c$ jet identification efficiency, as measured in dijet events along with the data $b$ jet identification efficiency. Two OPs are shown (a,b) the Loose and (c,d) Tight. The efficiencies are parameterized as a function of (a,c) $p_T$, for central jets and versus (b,d) $\eta$.} \label{fig:cTRFs}
\end{figure*}

\begin{figure*}\centering
\includegraphics[width=0.48\textwidth]{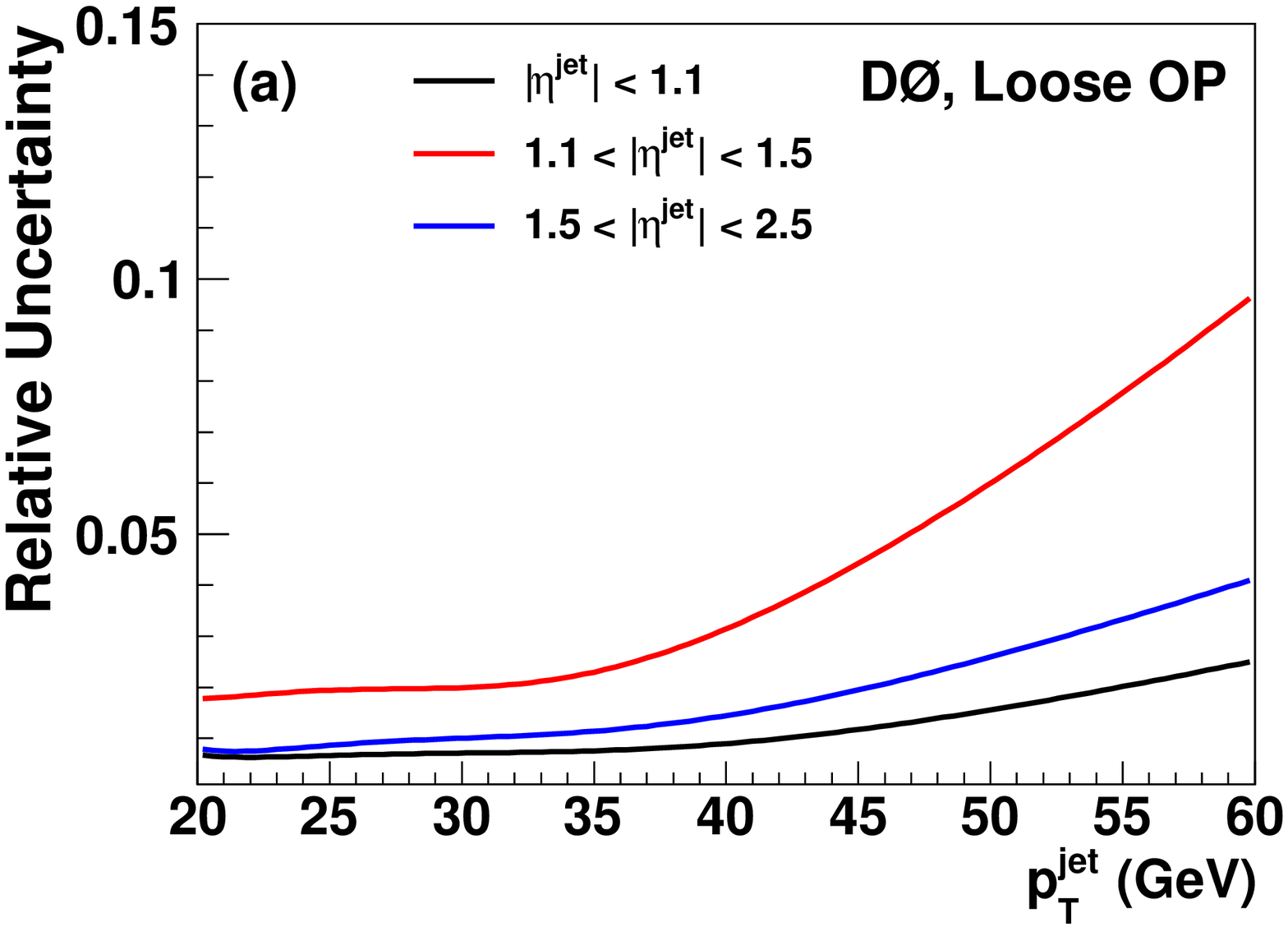} \includegraphics[width=0.48\textwidth]{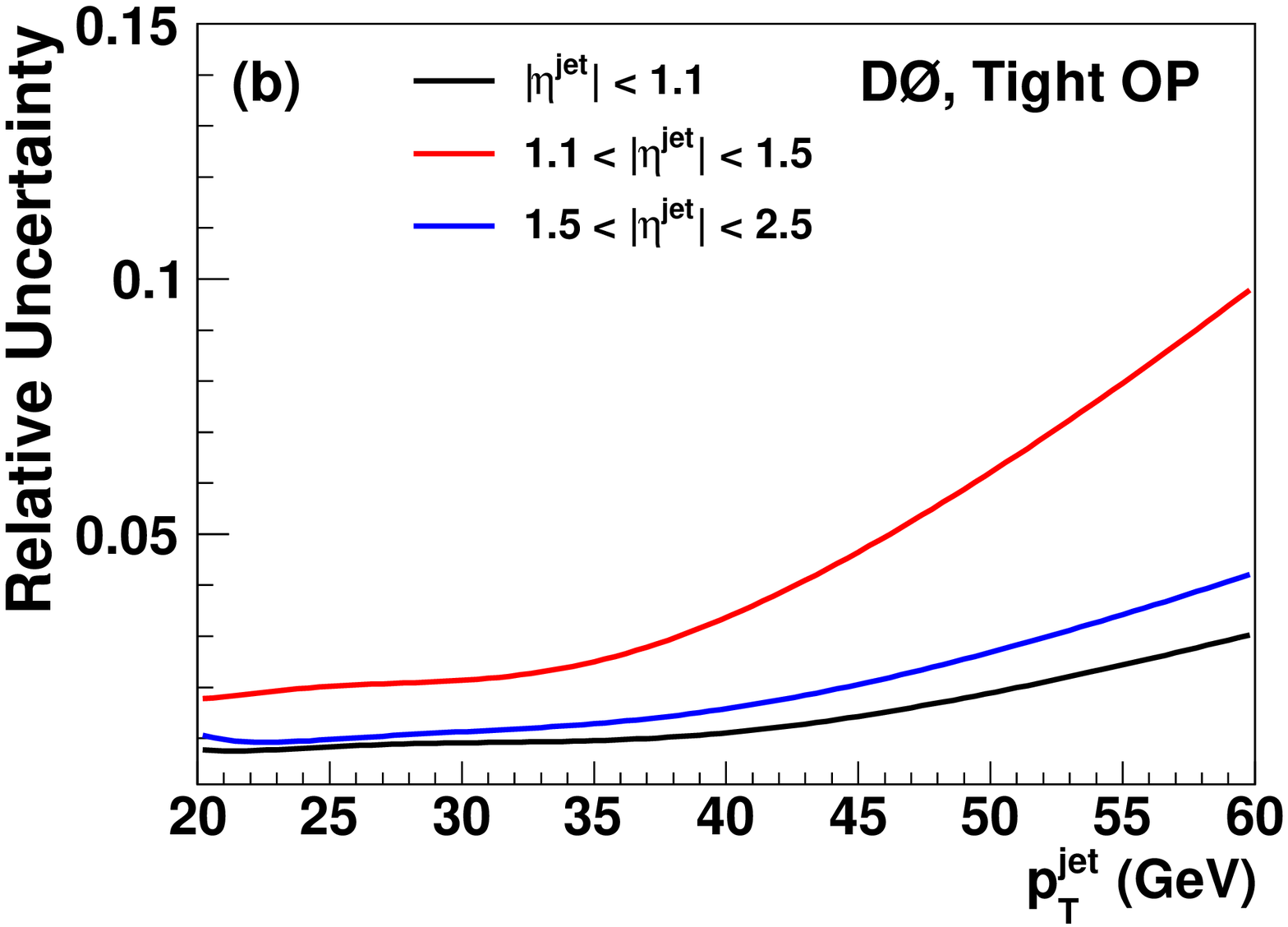}
\caption{(color online) The total uncertainty on $\varepsilon^{\text{data}}_{b}$ from the S8 method as a function of $p_T$~for two choices of OPs (a) Loose and (b) Tight.} \label{fig:totSyst}
\end{figure*}

In Fig.\thinspace\ref{fig:hSF} the efficiency for identifying a muonic $b$ jet, $\varepsilon_{b\rightarrow \mu X}$, is shown 
for data and MC. The ratio of these
two efficiencies, $SF$, is also displayed. 
Figs.\thinspace\ref{fig:bTRFs} and~\ref{fig:cTRFs} show the MC and data corrected 
efficiencies for $b$ and $c$ jets in dijet events, respectively. 
The data efficiency curves are corrected with the parameterized correction
 factor derived in Fig.\thinspace\ref{fig:hSF}.  
 Finally, in Fig.\thinspace\ref{fig:totSyst}, we present the total systematic uncertainty 
 for the S8 method on $\varepsilon^{data}_{b}$, discussed in Ref.\thinspace\cite{bid_nim}, parameterized as a function of jet $p_T$.

\section{Misidentification Rate Determination}
\label{sec:fake_rate}

A precise understanding of the misidentification rates is especially important 
in searches for rare processes which can be overwhelmed by large backgrounds.
Previous methods~\cite{bid_nim, cdfhobit, cmsbtag} to determine this 
rate relied heavily on simulation.
The method in Ref.\thinspace\cite{bid_nim} for estimating the misidentification 
rate uses ``negatively tagged" (NT) jets, or those with negative IP, with input from simulation.
Here we present the {\it SystemN} (SN) method which 
extracts misidentification rates directly from data.

\subsection{SystemN method}

The SN method uses a series of linear equations to describe the efficiency for light jets to satisfy the various \bl~OPs. 
Using a data sample of inclusive dijet events (the {\it inclusive jet} sample) we separate events as determined by the OP boundaries. 
If we have $n$ OPs, then there will be $n$+1 bins, with each bin
containing all the jets between the two consecutive OP's \bl~values. 
An equation relating the number of jets of each flavor, along with
their identification efficiencies, to the total number of retained jets in
each bin is formed:

\begin{eqnarray}
N &=& \varepsilon_{l}n_{l} + \varepsilon_{c} n_{c} + \varepsilon_{b} n_{b},
\end{eqnarray}

\noindent where $N$ is the number of selected jets in that bin,
$\varepsilon_{X}$ is the efficiency to identify a jet of flavor $X$, and
$n_{X}$ is the number of jets of flavor $X$ in the total sample.
The measured $b$ and $c$ jet efficiencies  from the S8 method are
used to predict the rate for selecting $b$ and $c$ jets in each bin.
For example, the equations describing a selection of five arbitrary 
OPs is given below (a total of twelve OPs are defined in the real analysis):

\begin{eqnarray}
\label{eq:SN}
  \begin{array}{llll}

\varepsilon^{OP5}_{l}n_{l} &+ \varepsilon^{OP5}_{c} n_{c} &+ \varepsilon^{OP5}_{b} n_{b} &= N_{OP5} \\
\varepsilon^{OP4-5}_{l}n_{l} &+ \varepsilon^{OP4-5}_{c} n_{c} &+ \varepsilon^{OP4-5}_{b} n_{b}  &= N_{OP4-5}\\
\varepsilon^{OP3-4}_{l}n_{l} &+ \varepsilon^{OP3-4}_{c} n_{c} &+ \varepsilon^{OP3-4}_{b} n_{b}  &= N_{OP3-4}\\
\varepsilon^{OP2-3}_{l}n_{l} &+ \varepsilon^{OP2-3}_{c} n_{c} &+ \varepsilon^{OP3-4}_{b} n_{b}  &= N_{OP2-3}\\
\varepsilon^{OP1-2}_{l}n_{l} &+ \varepsilon^{OP1-2}_{c} n_{c} &+ \varepsilon^{OP1-2}_{b} n_{b}  &= N_{OP1-2}\\
\varepsilon^{aOP1}_{l}n_{l} &+ (1-\varepsilon^{OP1}_{c}) n_{c} &+ (1-\varepsilon^{OP1}_{b}) n_{b} &= N_{OP1},
  \end{array}
\end{eqnarray}

\noindent where $\varepsilon^{OPi-j}_{X}$ is the efficiency for selecting a jet of flavor 
$X$ between the the $i^{th}$ and $j^{th}$ OP boundaries. The anti-OP1 point, aOP1, 
is the set of all jets which fall below the OP1 requirement. 
The number of jets of a given flavor, $n_X$, can be extracted from the data using 
a template fit based on the \msv~distributions corresponding to each jet flavor, as described below. 

\subsection{Sample composition} \label{sec:sc}

A measurement of the overall flavor composition is obtained by fitting \msv~templates 
for $b$, $c$, and light jets to a data distribution. 
These fits provide the number of $b$ and $c$ jets after the \bl~and SVT requirements, 
$n_b^{M_{SV}}$ and $n_c^{M_{SV}}$. 
Applying these requirements creates a sample enriched in heavy flavor jets.
The sample composition of the {\it inclusive jet} sample 
is calculated by extrapolating from this {\it heavy flavor} sample using $b$ and $c$ jet selection 
efficiencies measured using the S8 procedure for jets passing \bl~and SVT 
requirements. The data sample is divided into several jet \pt and $\eta$ bins to provide
a parameterization of the sample composition. 

Data is used to estimate the \msv~template shapes for the 
different jet flavors.  
For the $b$ and $c$ jet \msv~templates, a data-to-MC correction factor is 
estimated by comparing the \msv~distributions in a separate data sample (described in Sec.\thinspace\ref{sec:HFtemps}) 
to the MC templates on a bin-by-bin basis.
For light jets, \msv~template shapes are estimated using a data sample 
enriched in light jets, described in Sec.\thinspace\ref{sec:NTtemps}.

\subsubsection{Corrections to the heavy flavor templates}\label{sec:HFtemps}

To obtain an estimate of the shape of the heavy flavor jet \msv~distribution from data, a heavy flavor 
enriched dijet sample is constructed by requiring:

\begin{itemize}
\item Two taggable jets with a separation of $|\Delta \phi(\text{jet}_1, \text{jet}_2)| > 2.5$.
\item A jet must be selected by passing both an \bl~and SVT requirement.
\item The recoiling jet must be matched to a muon, $p_{T}^{\mu} > 8$~GeV, and pass a SVT requirement with \msv~$>1.8 $~GeV.
\end{itemize}

\begin{figure}\centering
\includegraphics[width=0.48\textwidth]{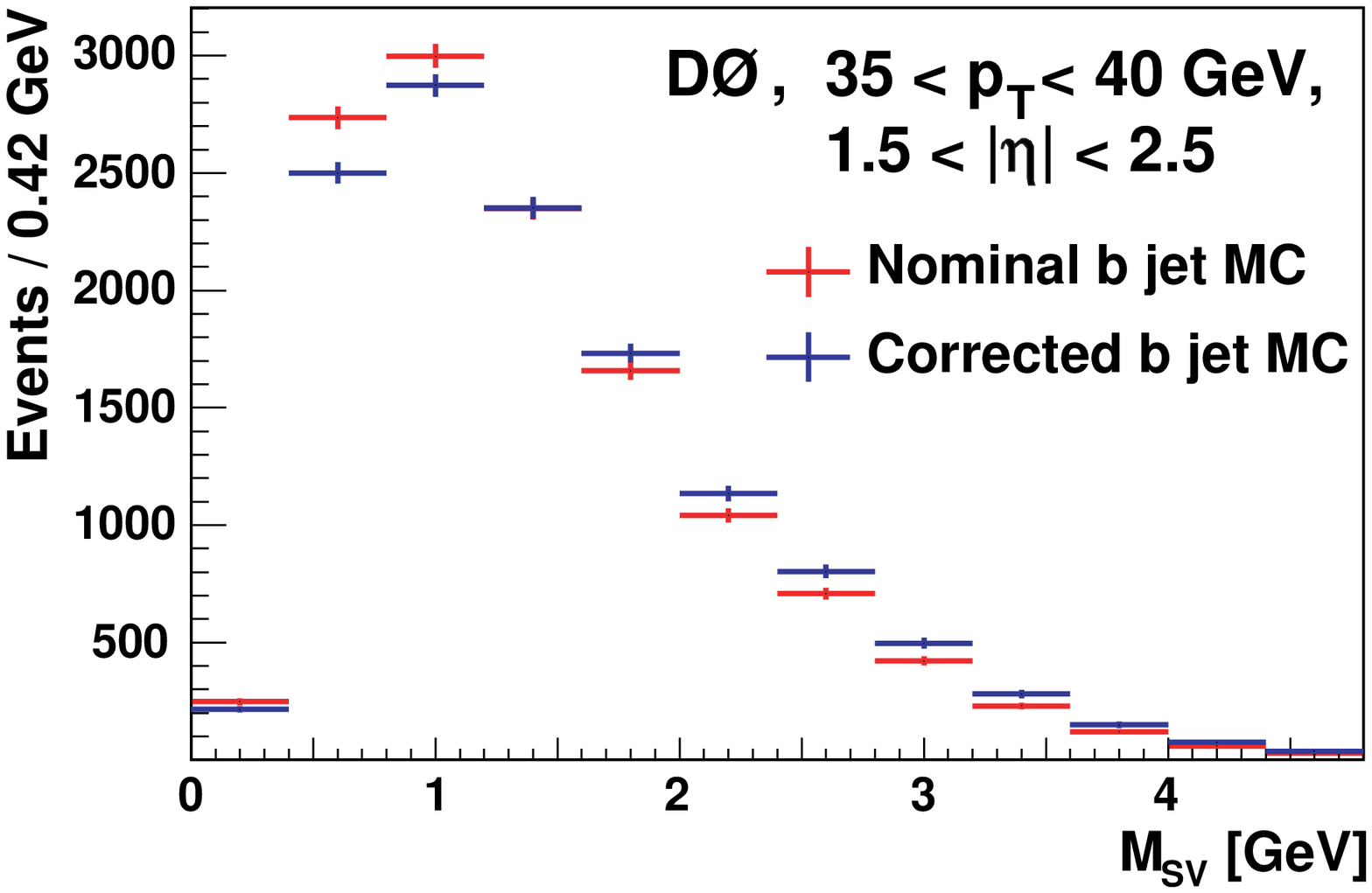}
\caption{(color online) Comparison of MC and data corrected $b$ jet \msv~template shapes for jets with $1.5 < |\eta| < 2.5$ and $35<p_{T}<45~\text{GeV}$. The data corrected \msv~template uses a shape reweighting derived in Sec.\thinspace\ref{sec:HFtemps}.} \label{fig:datacorrmc_bshape}
\end{figure}

The ratio of the data \msv~distribution and the MC predicted \msv~templates
for $b$, $c$, and light jets are used to create a correction factor. To determine the 
normalization of the MC templates the predicted sample composition is taken from the MC.
This correction factor is then applied to the MC $b$ and $c$ jet \msv~templates to correct their shape 
to the data in separate jet \pt and $\eta$ bins, 
an example of the corrected mass template is shown in Fig.\thinspace\ref{fig:datacorrmc_bshape}.

\subsubsection{Data driven light jet templates}\label{sec:NTtemps}

\begin{figure}\centering
\includegraphics[width=0.48\textwidth]{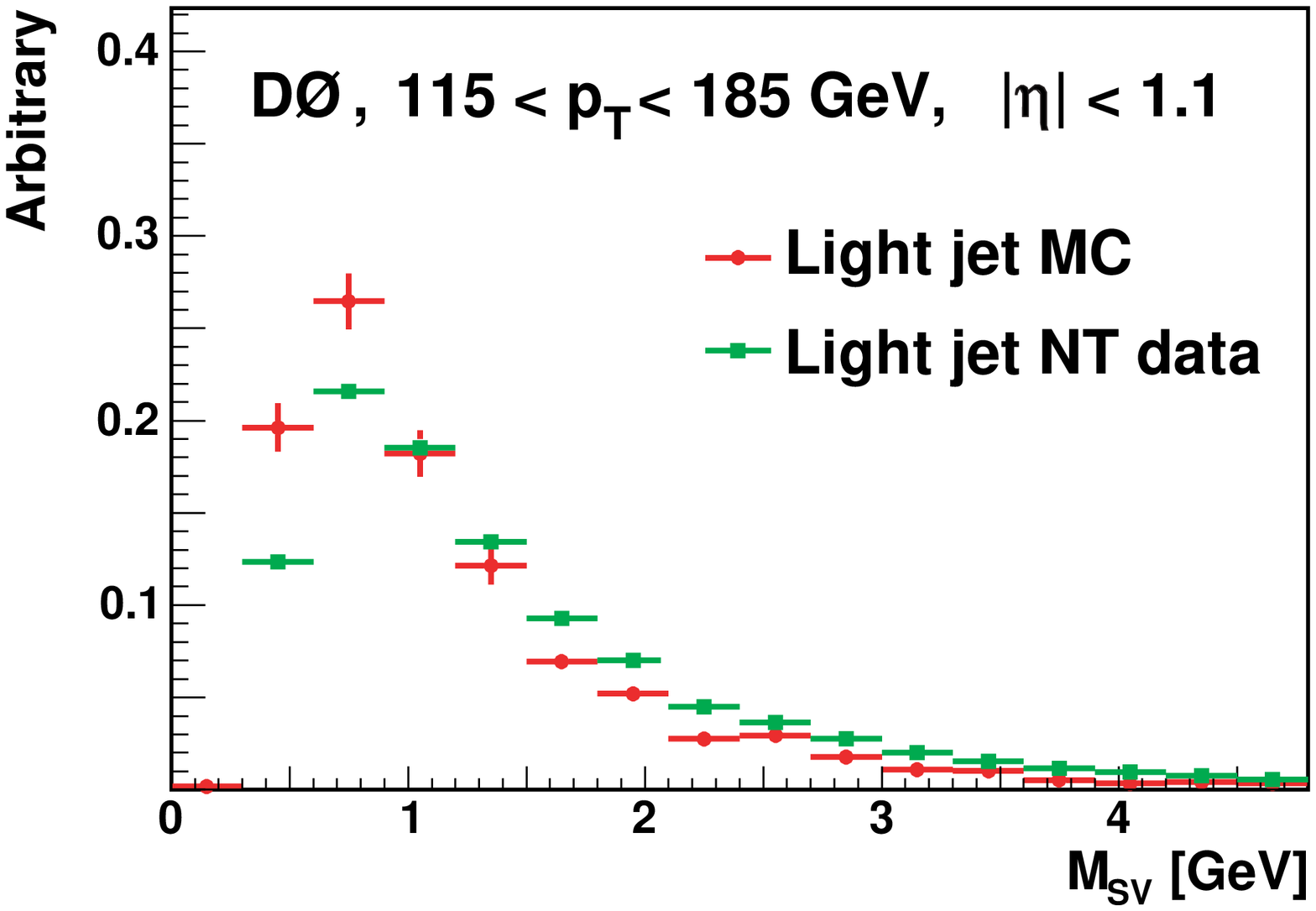}
\caption{(color online) Comparison of the MC light jet and NT \msv~mass templates for jets with $|\eta| < 1.1$ and $115 < p_T < 185$~GeV. The difference is taken as a measure of the systematic uncertainty due to residual contamination from 
heavy flavor jets in the NT data.}  \label{fig:ntshape_cc}
\end{figure}

The light jet templates are estimated from \msv~distribution of jets in a NT data sample~\cite{bid_nim}. 
This sample comprises jets having a negative IP and passing an SVT selection.
The shape of the \msv~distribution corresponding to this sample is affected by 
contamination due to the presence of heavy flavor jets and as such 
is not a perfect representation of the light jet \msv~shape in data. The NT template shapes 
are measured from data in each \pt and $\eta$ interval. 
Fig.\thinspace\ref{fig:ntshape_cc} shows a comparison between the NT \msv~distribution and the MC light jet template.
The difference in the shapes is taken as a systematic uncertainty.

\begin{figure}\centering
\includegraphics[width=0.48\textwidth]{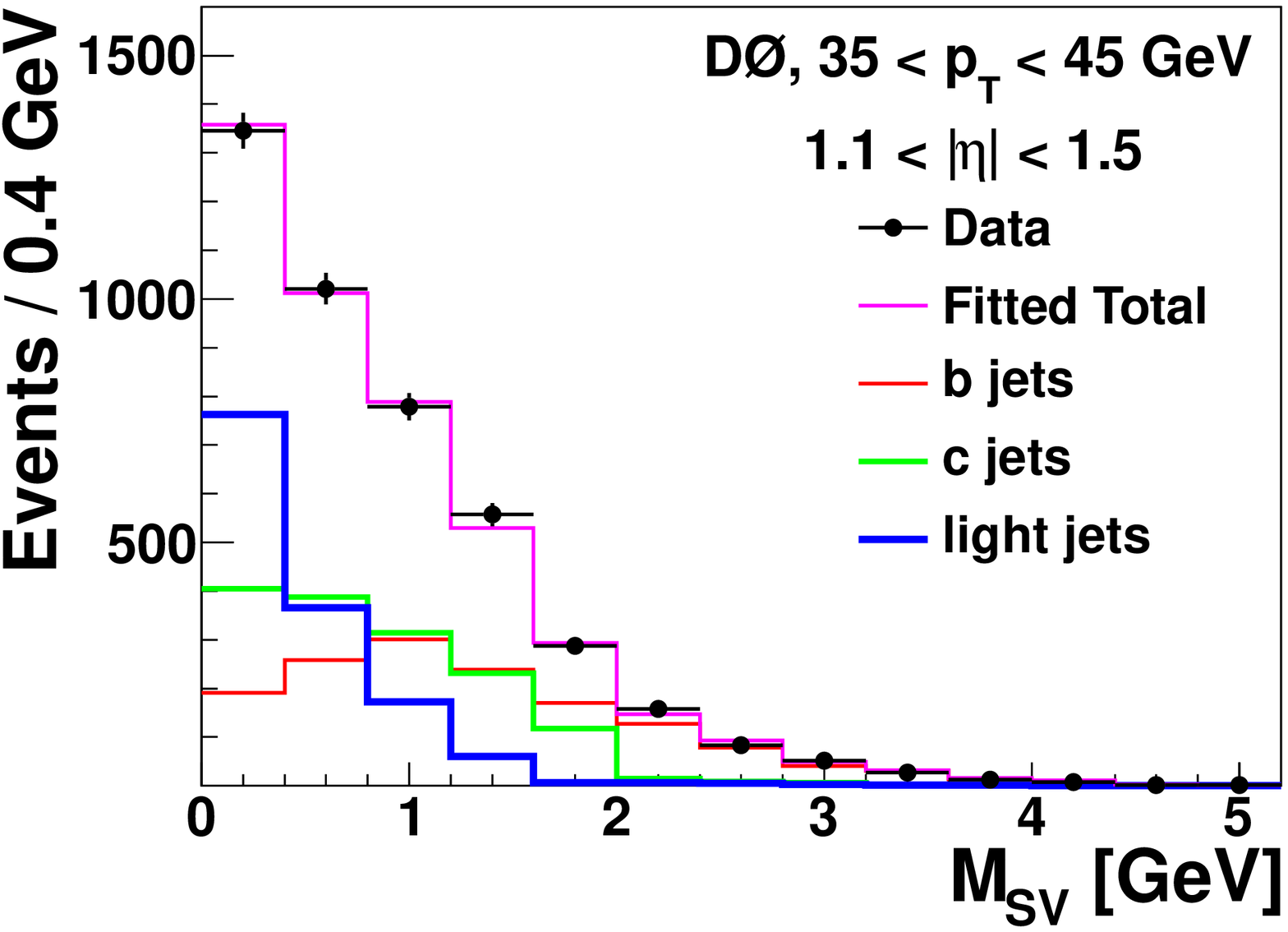}
\caption{(color online) An example of the sample composition fit using the \msv~for jets which 
pass \bl~and SVT requirements and have $35 < p_{T} < 45~\text{GeV}$ and $1.1 < |\eta| < 1.5$. 
The $b$, $c$, and light jets are fit to the data resulting in the total fitted distribution.} \label{fig:sc_fits_cc_I}
\end{figure}

\subsubsection{Sample composition measurement}

The data driven templates obtained above are used to fit the \msv~distribution in data using a log-likelihood fitter 
in bins of jet \pt and $\eta$. An example of a fit to the \msv~distribution 
using the $b$, $c$, and light jet templates is shown in Fig.\thinspace\ref{fig:sc_fits_cc_I}.
This results in a measurement of the fraction of each flavored jet type in that bin.
The fits in each of the \pt and $\eta$ regions 
are subsequently extrapolated back to the full {\it inclusive jet} sample using 
the $b$ and $c$ jet efficiency distributions measured for the \bl~and SVT 
algorithms. The number of events of heavy flavor, $HF$ (either $b$ or $c$),
in the {\it inclusive jet} sample is calculated using the following formula:

\begin{equation}
n_{HF} = N \times f_{HF} = N \times \frac{f_{HF}^{Tag}}{\varepsilon_{HF}^{Tag}}
\end{equation}

\noindent where $f_{HF}^{Tag}$ is the fraction of jets with flavor $HF$ extracted from the heavy flavor 
enriched sample and $\varepsilon_{HF}^{Tag}$ is the S8 efficiency for a \bl~and SVT requirements, and $N$ is the total
number of events in that bin. 
The efficiency is calculated for the average \pt and $\eta$ of the jets in the region. 
While $f_{HF}^{Tag}$ can be corrected to the {\it inclusive jet} sample, the light jet 
fraction cannot be. The corresponding light jet fraction in the {\it inclusive jet} sample 
is then determined from $f_l = 1-f_b-f_c$.

The parameterization of the {\it inclusive jet} sample composition is important to obtain 
the misidentification rate as a function of \pt and
to minimize the effect of statistically limited bins at high \pte. However, the choice of 
parameterization is not straightforward. 
The optimal parameterizations were determined by considering the $\chi^{2}$ probability of various functional forms, 
typically a first order polynomial or a second order logarithmic polynomial. 

\subsection{Solutions of the SystemN equations}

Instead of solving Eq.\thinspace\ref{eq:SN} analytically, we form a likelihood to improve the
stability of the solutions. In this likelihood 
we take the equations and compare them to what is predicted from simulations. 
We allow the extracted flavor fractions, $f_X$, to float within their uncertainties during 
this fit. To help constrain this likelihood a second set of SN equations is built 
using a new data sample, the full procedure 
is repeated and added to the likelihood fit. This new sample is a sub-set
of the {\it inclusive jet} sample which has the additional requirement that the recoiling ``away jet" must be matched to a muon. This
sample is defined as the ``{\it away} jet sample".

The resulting likelihood is formed by summing over each of the OP bins for both samples:

\begin{equation}
LLH = -2 \sum^{N_{S}}_{S}\sum^{N_{OP}}_{x=OP} (N^{S}_{x}\ln(N^{MC}_{x})-N^{MC}_{x})
\label{eqn:llh} \end{equation}

\noindent where $N^{S}_{x}$ is the number of data events in sample $S$, either {\it inclusive} or {\it away jet} sample,
in the \bl~interval $x$, $N^{MC}_{x}$ is the
predicted number of events in OP bin $x$. 
A normalization factor, $LLH_{Norm}$, is used to ensure that the likelihood values remain well defined:

\begin{equation}
LLH_{Norm} = -2 \sum^{N_{S}}_{S} \sum^{N_{OP}}_{x=OP} (N^{S}_{x} \ln(N^{S}_{x})-N^{S}_{x})
\end{equation}

\noindent which is then subtracted from the likelihood.

We use the $b$ and $c$ jet fractions measured in the previous section to help
stabilize the fit through a term which is added to the likelihood:

\begin{equation}
d^{T}E^{-1}d.
\end{equation}

\begin{figure*}\centering
\includegraphics[width=0.48\textwidth]{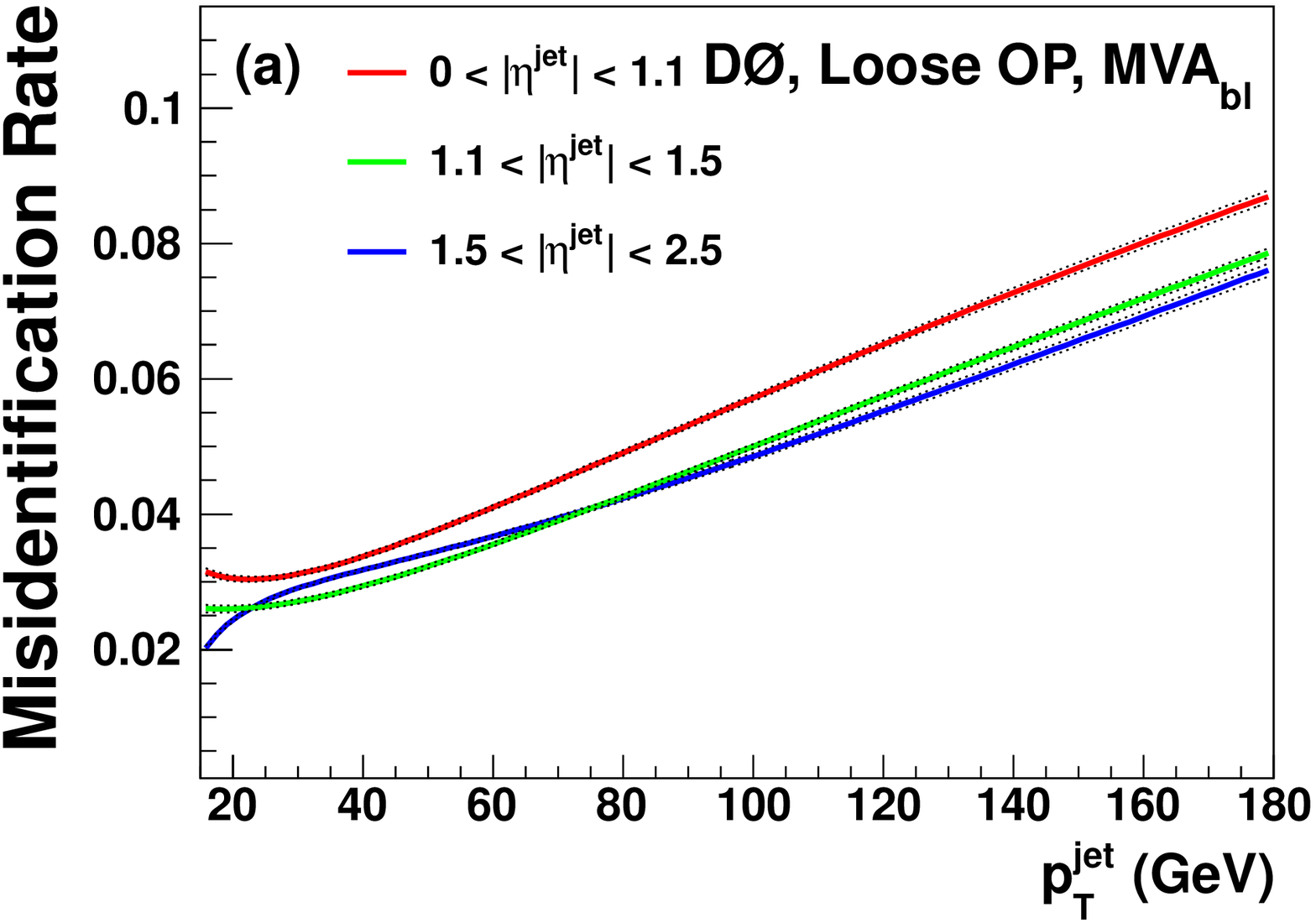} \includegraphics[width=0.48\textwidth]{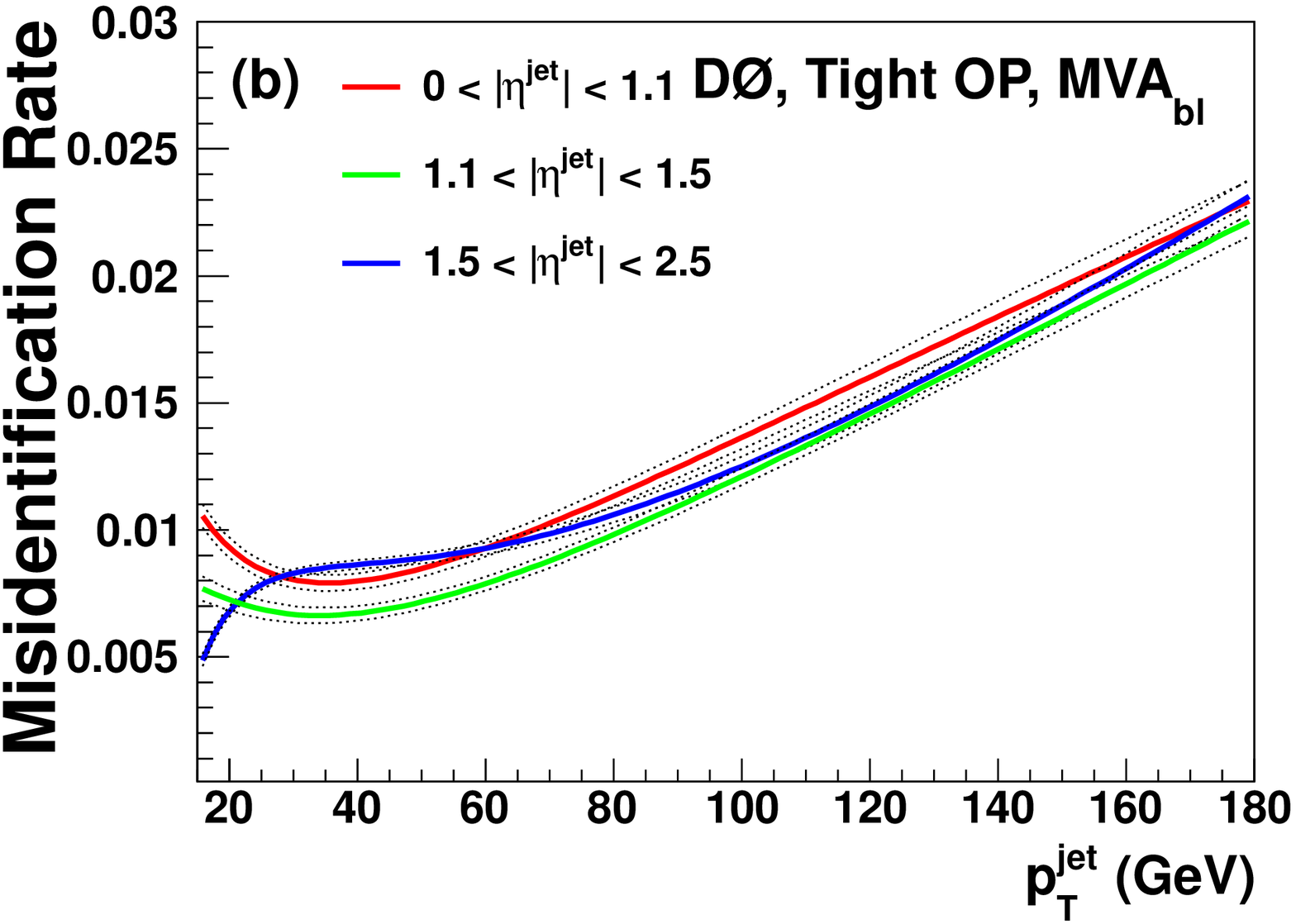}
\caption{(color online) The SN data driven misidentification rates for the \bl~algorithm. 
Two OPs are shown (a) Loose and (b) Tight. 
These are further parameterized over jet $p_T$~and for three different jet $\eta$ intervals: 
$0<|\eta|<1.1$, $1.1<|\eta|<1.5$, and $1.5<|\eta|<2.5$.
The black dotted lines represent the uncertainty on the fit.
} \label{fig:FRtrfs}
\end{figure*}

\noindent $E$ is a $2\times2$ covariance error matrix resulting from 
the extraction of the $b$ and $c$ jet content from the \msv~fit and $d$ is a vector
\begin{equation}
  d = \left(
    \begin{array}{c}
      n_b - n_b^{M_{SV}}\\
      n_c - n_c^{M_{SV}}
    \end{array}
  \right),
\end{equation}
\noindent where $n_{x}^{M_{SV}}$ is the number of jets, of flavor $x$, estimated from the \msv~template fits, 
and $n_{x}$ are the number of jets, of flavor $x$, in the {\it inclusive sample}.
The result of this likelihood fit is the extraction of the final data driven light jet efficiency parameterized over jet
\pt and $\eta$ in OP bins. These misidentification rates are shown in Fig.\thinspace\ref{fig:FRtrfs}.

\subsection{SystemN systematic uncertainties}

The three dominant systematic uncertainties on the misidentification rates are:

\begin{itemize}
\item The shape of the $b$ and $c$ jet \msv~templates
\item The shape of the light jet \msv~template
\item The uncertainty on the $b$ and $c$ jet efficiencies from the S8 method
\end{itemize}

\paragraph*{{\bf Heavy flavor template shape.}}

The effect of imperfections in the modeling of the $b$ and $c$ jet \msv~templates is 
estimated by carrying out the sample composition measurement using a set of 
heavy flavor \msv~templates which are not corrected to data in each of the \pt and 
$\eta$ intervals. The full difference between the MC and data corrected sample composition 
predictions is used as an uncertainty. As described in Sec.\thinspace\ref{sec:HFtemps}, the heavy flavor templates
are derived using MC inputs. These inputs are then varied and the largest deviation from the nominal shape 
is used to provide an additional uncertainty.

\paragraph*{{\bf Light flavor template shape.}}

The uncertainty due to the shape of the light jet \msv~templates is estimated by 
performing the sample composition fit using both the NT and MC light jet 
template shapes, taking the difference in the sample composition to assign 
an uncertainty. 

\paragraph*{{\bf {\boldmath $b$} and {\boldmath $c$} jet efficiency uncertainty}.}

When extrapolating the flavor fractions, measured in the heavy flavor enriched sample, 
to the {\it inclusive jet} sample the efficiencies from the S8 method are used. To account for the 
uncertainties inherited in this procedure it is repeated after the efficiencies are varied by 
$\pm1\sigma$. This variation will only affect the extrapolation procedure.  

\begin{figure*}\centering
\includegraphics[width=0.48\textwidth]{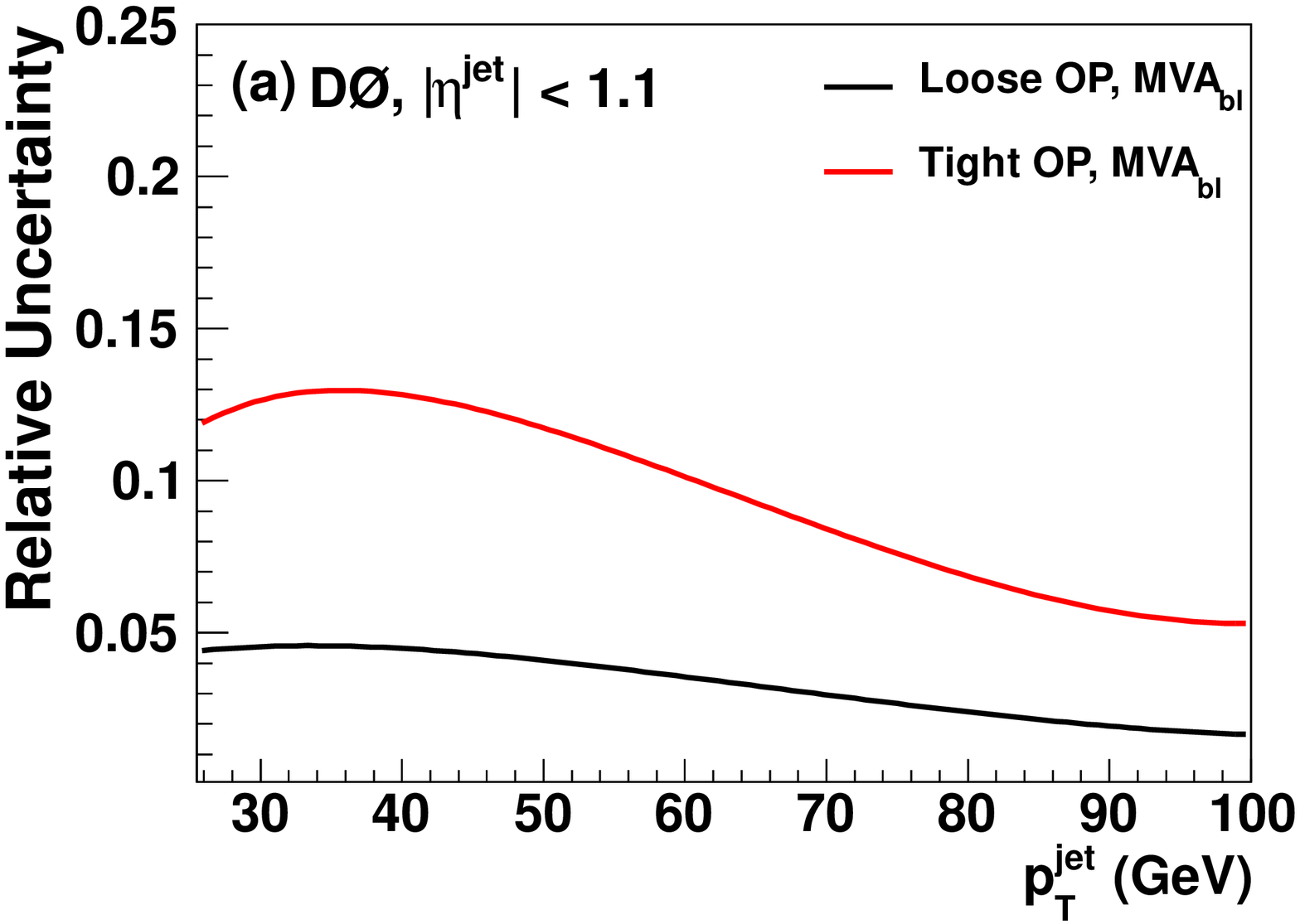}\includegraphics[width=0.48\textwidth]{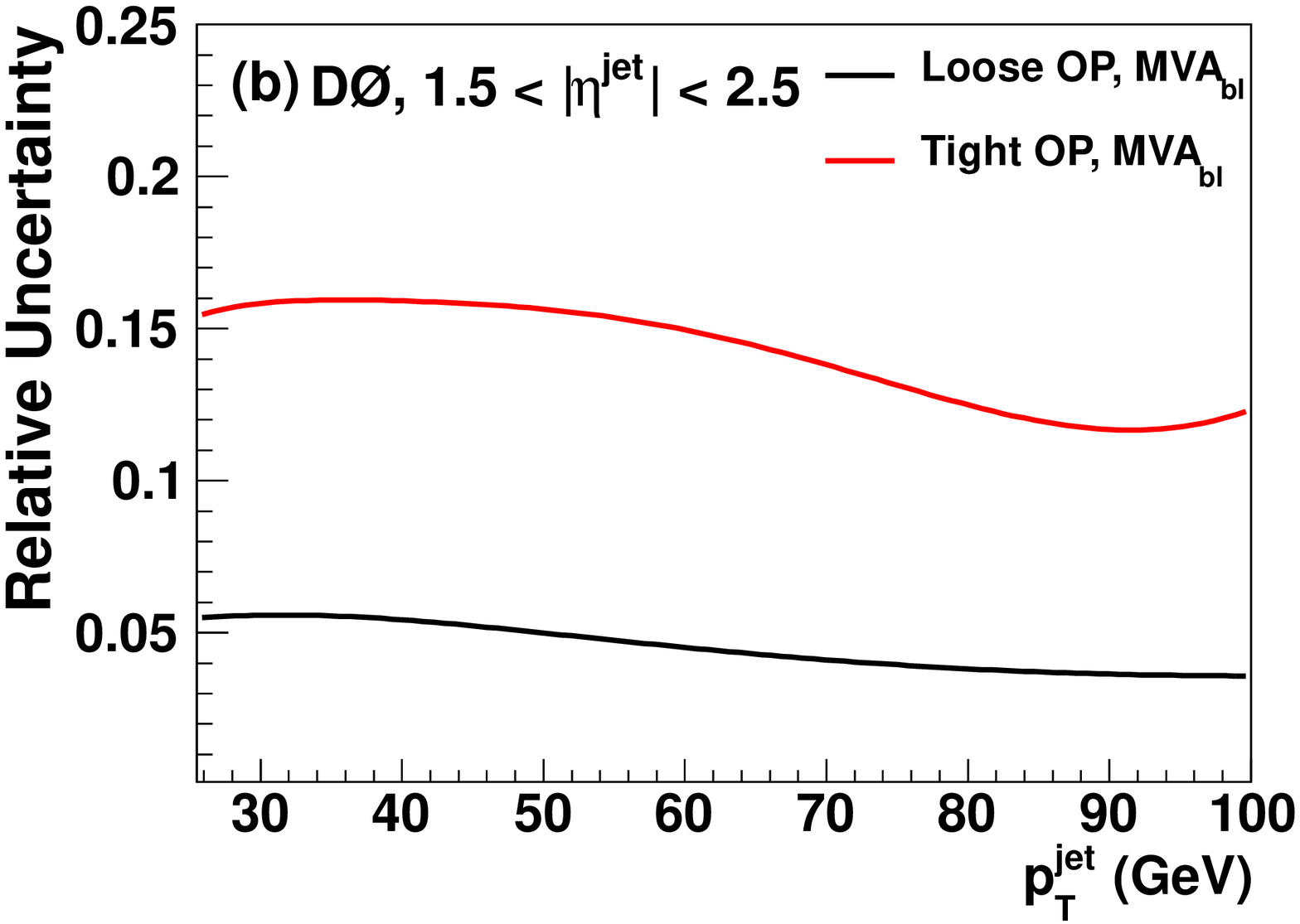}
\caption{(color online) The total relative uncertainty on the misidentification rate from the SN method 
parameterized in terms of jet $p_T$~and for two different $\eta$ regions: 
(a) $|\eta| <1.1$ and (b) $1.5<|\eta| < 2.5$.} \label{fig:etaerror_fit}
\end{figure*}

The parameterization of the systematic uncertainties is evaluated by carrying out closure tests, where
the percentage difference between the number of actually
selected jets and the predicted number of jets in various
bins in  $p_{T}$ and $\eta$ regions are compared. The uncertainty
is determined from the RMS of the resulting distributions. 
The total uncertainty on the data-driven misidentification rate attained 
using the SN method, given by the statistical and systematic uncertainties 
combined in quadrature, is shown in Fig.\thinspace\ref{fig:etaerror_fit} for the Loose and Tight 
OPs of the \bl~algorithm.

\subsection{Comparison with previous method}

\begin{figure*}\centering
\includegraphics[width=0.48\textwidth]{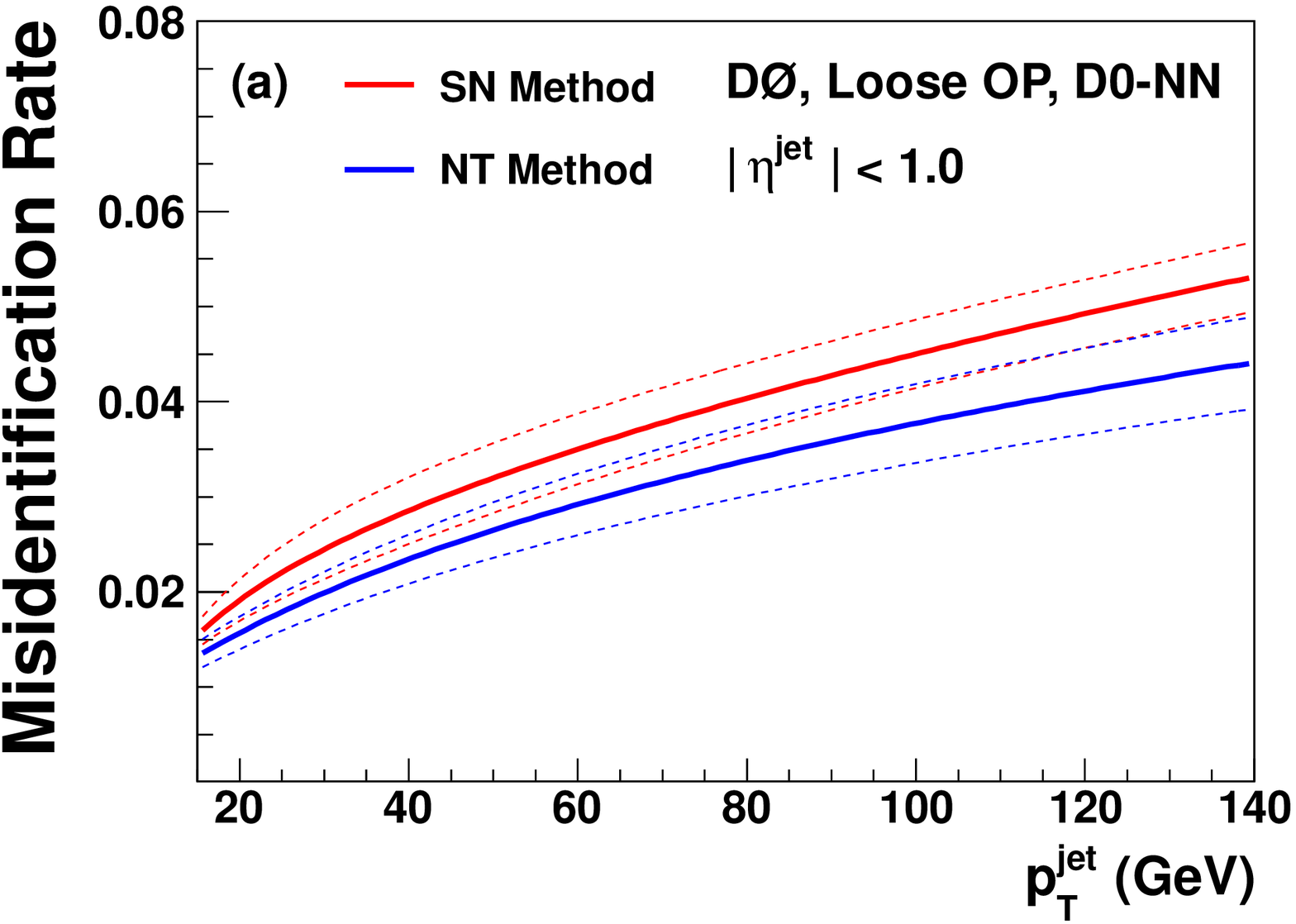}\includegraphics[width=0.48\textwidth]{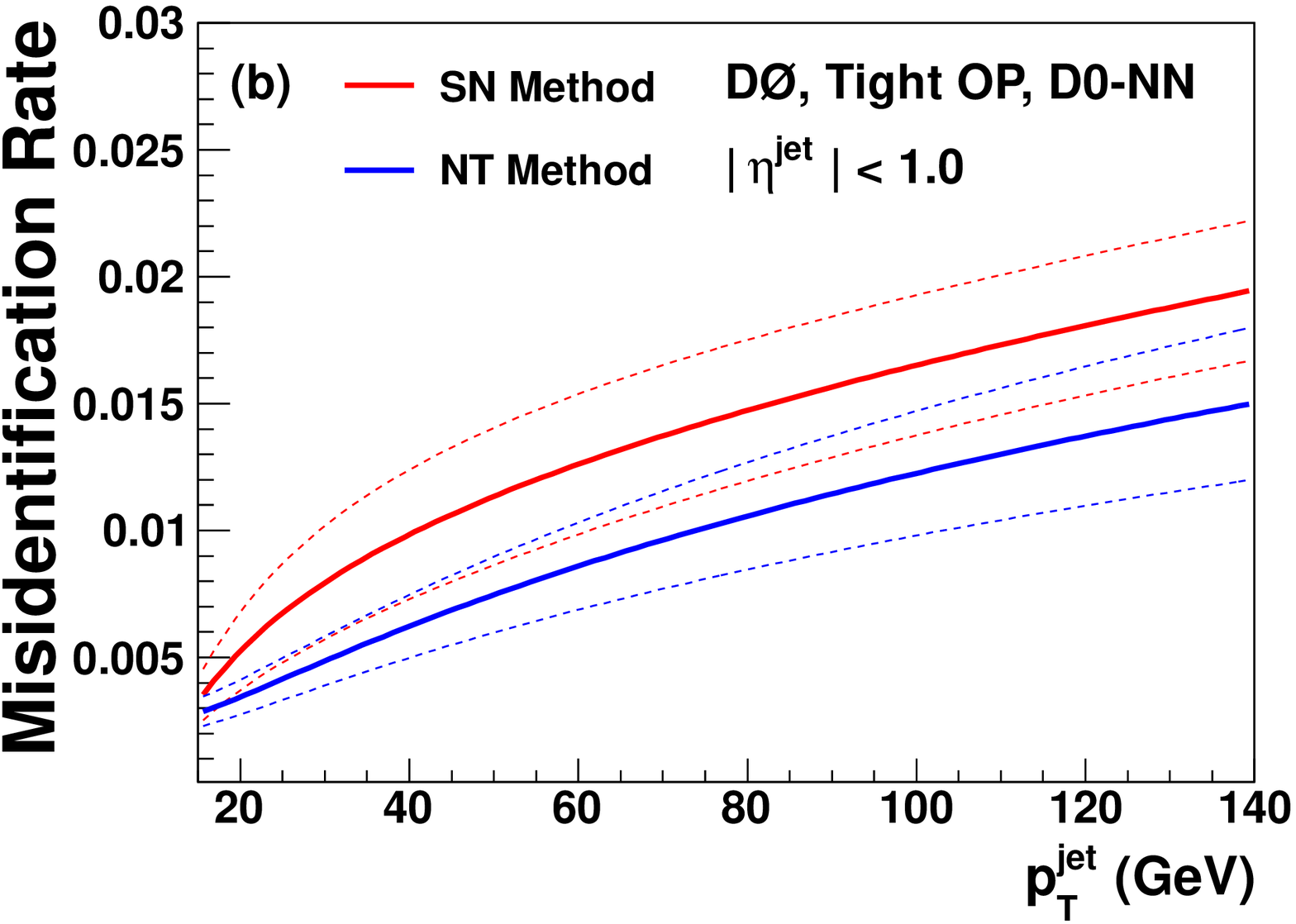}
\caption{(color online) Comparison between the misidentification rates of the 
D0-NN derived for two OPs, (a) Loose and (b) Tight, using the new SN method 
and the old method described in Ref.\thinspace\cite{bid_nim}.
The dashed bands which surround the values correspond to the total uncertainties.} \label{fig:compVeryTight}
\end{figure*}

A comparison between the misidentification rates of the D0-NN algorithm measured using the SN
method and those estimated by the NT method of Ref.~\cite{bid_nim} is shown in
Fig.\thinspace\ref{fig:compVeryTight}. Both provide comparable uncertainties. 
For the looser OPs the central value of the new method gives
a misidentification rate roughly $20\%$ higher than the central values for the previous method, 
and for the tighter OPs the difference is closer to $35\%$. 
The two methods do agree with each other within uncertainties across the full range
of jet \pte, but the misidentification rate for the NT method is systematically lower.

The source of this difference comes from the use of simulation in the NT method. 
With the removal of the $V^0$s the main source of misidentified light jets comes from
detector resolution and track mis-reconstruction effects.
The simulation does not accurately reproduce these effects by modeling ideal detector responses
and the resulting misidentification rate as determined by the NT method is systematically underestimated.   

\subsection{\bl~misidentification rates}

\begin{figure*}\centering
\includegraphics[width=0.48\textwidth]{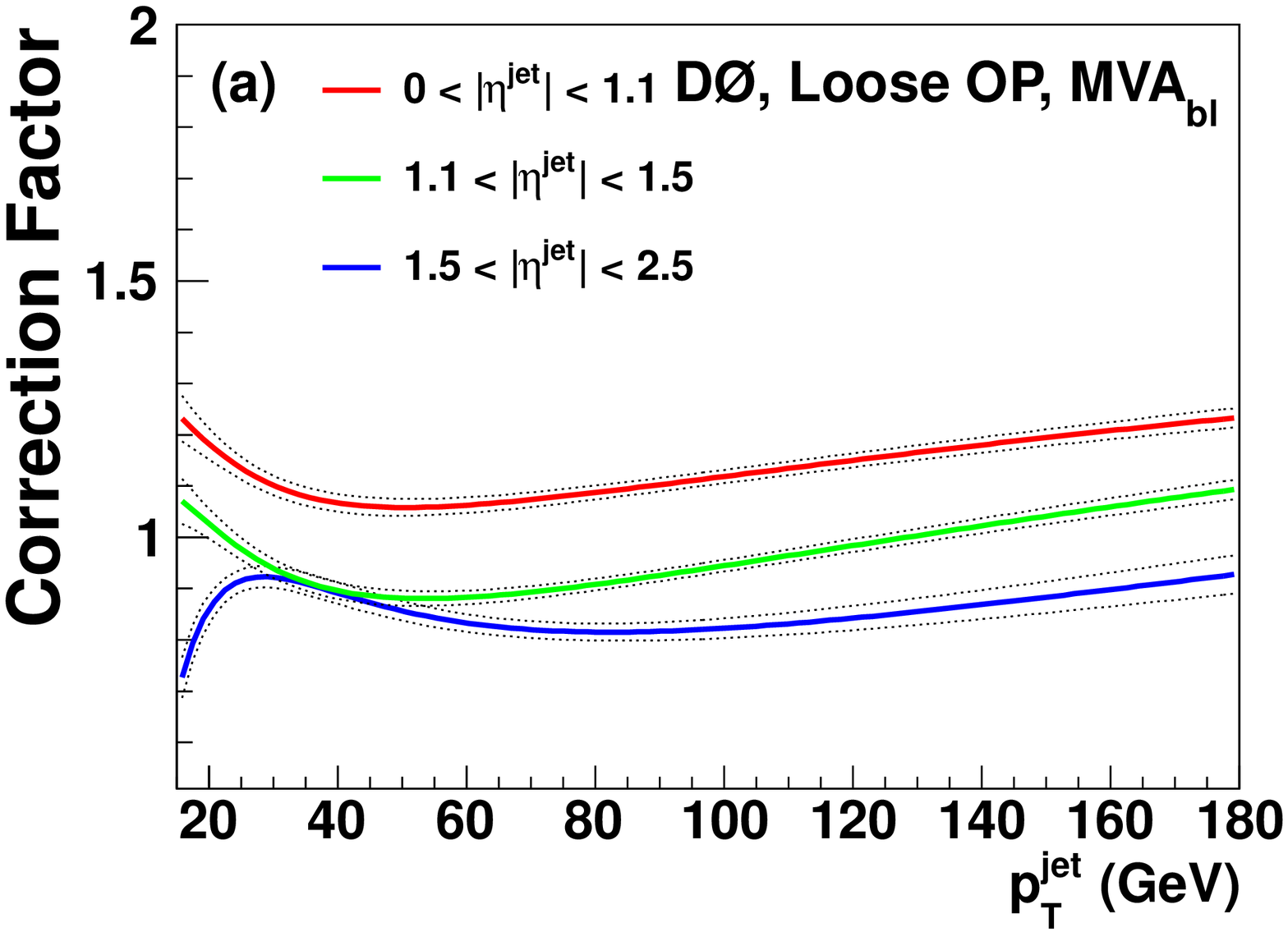} \includegraphics[width=0.48\textwidth]{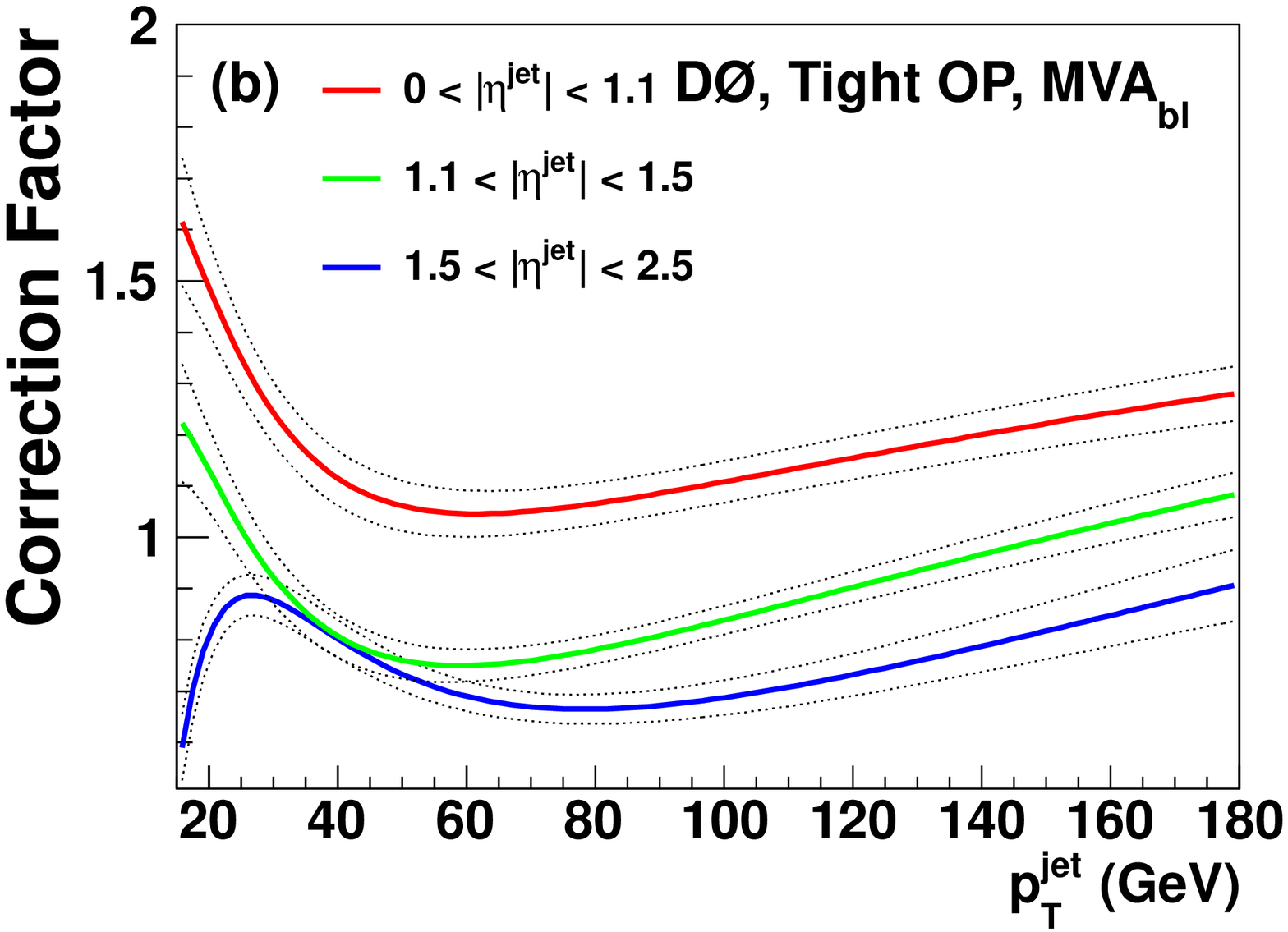}
\caption{(color online) The misidentification rate correction factors for the light jet MC which are derived by taking the ratio 
of the data and MC misidentification rates. Two OPs are shown, (a) Loose and (b) Tight. 
These are further parameterized over jet $p_T$~and for three different jet $\eta$ intervals: 
$0<|\eta|<1.1$, $1.1<|\eta|<1.5$, and $1.5<|\eta|<2.5$.
The black dotted lines represent the uncertainty on the fit.} \label{fig:FRSF}
\end{figure*}

The final results are the misidentification rate for light jets extracted from our data, as shown in Fig.\thinspace\ref{fig:FRtrfs}. 
These are parameterized in terms of $p_T$~for three different $\eta$ regions. 
This data-driven measurement of the misidentification rate can be combined with that 
modeled in simulation and we can derive a MC correction factor, as shown in Fig.\thinspace\ref{fig:FRSF}. 
These correction factors are applied in the light jet simulations (for jets passing the \bl~requirements).
Table\thinspace\ref{final} shows the responses, efficiencies, and misidentification rates, of the \bl~algorithm as measured in data.

\begin{table*}
\caption{The efficiency of selecting a $b$, $c$, or light jet using the \bl~as determined
by using the S8 and SN method directly from data for 12 OPs.
The total uncertainties are included along with the OP definitions.}\label{final}
\begin{tabular}{|l|c|ccc|ccc|ccc|}
\hline 
\multirow{2}{*}{OP Name} & Min. & \multicolumn{3}{c|}{$\left|\eta^{\text{jet}}\right|<1.1$} & \multicolumn{3}{c|}{$1.1<\left|\eta^{\text{jet}}\right|<1.5$} & \multicolumn{3}{c|}{$1.5<\left|\eta^{\text{jet}}\right|<2.5$}\tabularnewline
\cline{3-11} 
 & \bl & $\varepsilon_{b}^{\text{data}}[\%]$ & $\varepsilon_{c}^{\text{data}}[\%]$ & $\varepsilon_{l}^{\text{data}}[\%]$ & $\varepsilon_{b}^{\text{data}}[\%]$ & $\varepsilon_{c}^{\text{data}}[\%]$ & $\varepsilon_{l}^{\text{data}}[\%]$ & $\varepsilon_{b}^{\text{data}}[\%]$ & $\varepsilon_{c}^{\text{data}}[\%]$ & $\varepsilon_{l}^{\text{data}}[\%]$\tabularnewline
\hline 
\hline 
L6 & 0.02 & $74.8\pm0.6$ & $39.2\pm0.3$ & $15.5\pm0.3$ & $75.3\pm0.6$ & $38.2\pm0.3$ & $13.9\pm0.5$ & $64.8\pm0.7$ & $31.9\pm0.4$ & $13.7\pm0.4$\tabularnewline
L5 & 0.025 & $73.2\pm0.6$ & $36.8\pm0.3$ & $13.3\pm0.3$ & $73.7\pm0.6$ & $36.0\pm0.3$ & $11.9\pm0.5$ & $62.7\pm0.7$ & $29.6\pm0.4$ & $11.6\pm0.2$\tabularnewline
L4 & 0.035 & $70.2\pm0.6$ & $33.0\pm0.3$ & $10.5\pm0.2$ & $70.7\pm0.7$ & $32.2\pm0.3$ & $9.4\pm0.5$ & $59.1\pm0.8$ & $25.8\pm0.3$ & $9.1\pm0.4$\tabularnewline
L3 & 0.042 & $68.9\pm0.7$ & $31.2\pm0.3$ & $9.2\pm0.2$ & $69.3\pm0.7$ & $30.4\pm0.3$ & $8.2\pm0.5$ & $57.5\pm0.8$ & $24.3\pm0.3$ & $8.0\pm0.4$\tabularnewline
L2 & 0.05 & $67.5\pm0.8$ & $29.6\pm0.3$ & $8.0\pm0.2$ & $68.0\pm0.8$ & $28.9\pm0.3$ & $7.2\pm0.5$ & $56.1\pm0.8$ & $23.0\pm0.3$ & $7.0\pm0.4$\tabularnewline
Loose & 0.075 & $63.8\pm0.8$ & $25.4\pm0.3$ & $5.63\pm0.2$ & $64.3\pm0.8$ & $24.9\pm0.3$ & $5.0\pm0.5$ & $51.9\pm1.0$ & $19.4\pm0.3$ & $4.9\pm0.2$\tabularnewline
oldLoose & 0.1 & $61.1\pm0.7$ & $22.8\pm0.3$ & $4.2\pm0.2$ & $61.7\pm0.7$ & $22.3\pm0.3$ & $3.8\pm0.5$ & $49.0\pm0.8$ & $17.2\pm0.3$ & $3.7\pm0.1$\tabularnewline
Medium & 0.15 & $56.7\pm0.6$ & $19.0\pm0.2$ & $2.6\pm0.2$ & $57.4\pm0.6$ & $18.7\pm0.2$ & $2.3\pm0.5$ & $44.5\pm0.8$ & $14.1\pm0.2$ & $2.4\pm0.4$\tabularnewline
Tight & 0.225 & $51.6\pm0.7$ & $15.4\pm0.2$ & $1.4\pm0.1$ & $52.4\pm0.7$ & $15.3\pm0.2$ & $1.3\pm0.4$ & $39.5\pm0.7$ & $11.1\pm0.2$ & $1.3\pm0.4$\tabularnewline
VeryTight & 0.3 & $47.4\pm0.6$ & $12.9\pm0.2$ & $0.9\pm0.2$ & $48.3\pm0.6$ & $12.9\pm0.2$ & $0.8\pm0.4$ & $35.4\pm0.7$ & $9.1\pm0.2$ & $0.8\pm0.4$\tabularnewline
UltraTight & 0.4 & $43.4\pm0.7$ & $10.9\pm0.2$ & $0.6\pm0.2$ & $44.8\pm0.6$ & $11.0\pm0.2$ & $0.4\pm0.4$ & $31.9\pm0.6$ & $7.4\pm0.2$ & $0.5\pm0.3$\tabularnewline
MegaTight & 0.5 & $40.4\pm0.6$ & $9.5\pm0.1$ & $0.4\pm0.2$ & $41.8\pm0.6$ & $9.6\pm0.1$ & $0.2\pm0.3$ & $29.2\pm0.7$ & $6.3\pm0.2$ & $0.4\pm0.3$\tabularnewline
\hline 
\end{tabular}
\end{table*}

\section{Summary and Conclusions}
\label{sec:conclusion}

The identification of heavy flavor jets is a crucial component of particle physics analyses. 
Utilizing the unique characteristics of the fragmenting $b$ quark we created algorithms 
which allow for the identification of $b$ jets with high efficiency and purity. 
The \bl~algorithm shows improvements over previous algorithms utilized at D0. 
For a light jet misidentification rate of 1\% we observe an 
improvement in the efficiency over the D0-NN algorithm for selecting a $b$ jet of 15\% per jet. 
A new method for extracting the misidentification rate directly from data has also been presented. 
The data-derived misidentification rates of the SystemN method are compatible within uncertainties 
with previous simulation-based methods, however a systematic difference is observed. This difference is due 
to the limited ability of the simulation to accurately model resolution and track mis-reconstruction effects. 
By removing this dependence on simulation the SystemN method provides a
more accurate and reliable measurement of the light jet misidentification rates in data.

\section*{Acknowledgement}
%
We thank the staffs at Fermilab and collaborating institutions,
and acknowledge support from the
DOE and NSF (USA);
CEA and CNRS/IN2P3 (France);
MON, NRC KI and RFBR (Russia);
CNPq, FAPERJ, FAPESP and FUNDUNESP (Brazil);
DAE and DST (India);
Colciencias (Colombia);
CONACyT (Mexico);
NRF (Korea);
FOM (The Netherlands);
STFC and the Royal Society (United Kingdom);
MSMT and GACR (Czech Republic);
BMBF and DFG (Germany);
SFI (Ireland);
The Swedish Research Council (Sweden);
and
CAS and CNSF (China).

\bibliographystyle{h-physrev3.bst}
\bibliography{NIM_v2.9}

\end{document}